\documentclass[12pt,preprint]{aastex}
\usepackage{emulateapj5,apjfonts}
\usepackage{natbib}
\usepackage{graphics}
\usepackage{amssymb}
\usepackage{onecolfloat}
\usepackage{bm}
\lefthead{Heitmann, White, Wagner, Habib, Higdon}
\righthead{Precision Determination of the Nonlinear Matter Power Spectrum}

\begin{document}


\def\head{
  \vbox to 0pt{\vss
                    \hbox to 0pt{\hskip 440pt\rm LA-UR-08-05630\hss}
                   \vskip 25pt}

\title{The Coyote Universe I: Precision Determination of the 
Nonlinear Matter Power Spectrum}
\author{Katrin~Heitmann\altaffilmark{1}, Martin White\altaffilmark{2}, 
        Christian Wagner\altaffilmark{3},
        Salman~Habib\altaffilmark{4}, and David
        Higdon\altaffilmark{5}}

\affil{$^1$ ISR-1, ISR Division,  Los
Alamos National Laboratory, Los Alamos, NM 87545}
\affil{$^2$ Departments of Physics and Astronomy, University of
California, Berkeley, CA 94720}
\affil{$^3$ Astrophysikalisches Institut Potsdam (AIP),
An der Sternwarte 16, D-14482 Potsdam}
\affil{$^4$ T-2, Theoretical Division, Los
Alamos National Laboratory, Los Alamos, NM 87545}
\affil{$^5$ CCS-6, CCS Division, Los Alamos National Laboratory, 
Los Alamos, NM 87545}

\date{today}

\begin{abstract}
  Near-future cosmological observations targeted at investigations of
  dark energy pose stringent requirements on the accuracy of
  theoretical predictions for the nonlinear clustering of matter.
  Currently, $N$-body simulations comprise the only viable approach to
  this problem. In this paper we study various sources of
  computational error and methods to control them. By applying our
  methodology to a large suite of cosmological simulations we show
  that results for the (gravity-only) nonlinear matter power spectrum
  can be obtained at 1\% accuracy out to $k\sim 1\,h\,{\rm Mpc}^{-1}$.
  The key components of these high accuracy simulations are: precise
  initial conditions, very large simulation volumes, sufficient mass
  resolution, and accurate time stepping. This paper is the first in a
  series of three; the final aim is a high-accuracy prediction scheme 
  for the nonlinear matter power spectrum that improves current
  fitting formulae by an order of magnitude.
\end{abstract}

\keywords{methods: $N$-body simulations ---
          cosmology: large-scale structure of universe}}

\twocolumn[\head]
\section{Introduction}

The nature of the dark energy believed to be causing the current
accelerated expansion of the Universe is one of the greatest puzzles
in the physical sciences, with deep implications for our understanding
of the Universe and fundamental physics. The twin aims of better
characterizing and further understanding the nature of dark energy are
widely recognized as key science goals for the next decade. Although
dark energy remains very poorly understood, theory nevertheless plays
an essential role in furthering this enterprise.

The phenomenology of cosmological models is theory-driven not only in
terms of providing explanations for the diverse phenomena that are
observed, as well as promoting alternative explanations of existing
measurements, but also due to the increasing reliance on theorists to
produce sophisticated numerical models of the Universe which can be
used to refine and calibrate experimental probes. Without a dedicated
effort to develop the tools and skill-sets necessary for the
interpretation of the next generation of experiments, we risk being
``theory limited'' in essentially all areas of dark energy studies.

As a concrete example of this general trend, forecasts for determination
of the dark energy equation of state and other cosmological parameters from
next-generation observations of cosmological structure typically assume
calibration against simulations accurate to the level of 1\% or better.
This target has rarely been met for simulations of complex nonlinear
phenomena such as the formation of large-scale structure in the Universe.
However it is precisely these probes, which provide information on both
the geometry of space-time and the growth of large-scale structure, which
will be key to unraveling the mystery of dark energy.

For upcoming measurements to be exploited to the full, theory must
reach not only the levels of accuracy justified by the measurements
but also cover a sufficiently wide range of cosmologies.  The problem
breaks down to two questions: (i) What is a reasonable coverage of
cosmological parameters, given the expected set of observations?  (ii)
What is the required accuracy for theoretical predictions -- over this
range of parameters -- for the given set of observations?  
It is crucial to realize that the ultimate requirement is on
controlling the {\em absolute\/} error -- taking into account all of the
relevant physics: gravity, hydrodynamics, and feedback mechanisms.
This is much more difficult to achieve than {\it relative\/} error control --
e.g.~asking what the relative importance of baryonic physics is versus
a baseline gravity-only simulation.
Most recent papers discuss the latter, implicitly assuming the existence
of a reference spectrum.  One aim of our work is to provide just such a
reference spectrum within the boundaries outlined.
We fully expect that the answers to both (i) and
(ii) will evolve, requiring more accurate modeling of a smaller range
of models, so we are most interested here in the near-term needs.
Associated with the first problem is the fact that, given the
impossibility of running complex simulations over the many thousands
of cosmologies necessary for grid based or Markov Chain Monte Carlo
(MCMC) estimation of cosmological parameters, one must develop
efficient interpolation methods for theoretical predictions.  These
methods must of course also satisfy the accuracy requirements of
question (ii).

The control of errors in the underlying theory for the CMB is adequate
to analyze results from Planck \citep{CMBcompare,Won08}. This is,
however, not the case for predictions of gravitational clustering in
the nonlinear regime, as is required for cluster counts, redshift
space distortions, baryon acoustic oscillations (BAO) and weak lensing
(WL) observations. In the case of BAO, the galaxy power spectrum in
the quasi-linear regime should be known to sub-percent accuracy, and
for WL the same is true for the mass power spectrum to significantly
smaller scales. Perturbation theory has errors on the mass power
spectrum currently estimated to be at the percent level in the weakly
nonlinear regime (see, e.g., \citealt{JeoKom06} and \citealt{Car09}
for recent treatments, or \citealt{Ber02} for an earlier review). To
reduce these errors, test the approximations, and model galaxy bias,
numerical simulations are unavoidable. Theoretical templates, in terms
of current power spectrum fits based on simulations (with errors at
the 5\% level), are already a limiting factor for WL observations at
wavenumbers $k\sim 1\,h\,{\rm Mpc}^{-1}$. \citet{HutTak05} show that
in order to avoid errors from imprecise theoretical templates
mimicking the effect of cosmological parameter variations, the power
spectrum has to be calibrated at about 0.5-1\% for $0.1\,h\,{\rm
  Mpc}^{-1}\le k \le 10\,h\,{\rm Mpc}^{-1}$. The scale most sensitive
for WL measurements is around $k\sim 1\,h\,$Mpc$^{-1}$ and $z\sim 0.5$
and the power spectrum therefore needs to be calibrated the most
accurately at that point (see, e.g, \citealt{HutTak05}, Figure~1). In a
very recent paper, \citet{Hil08} re-emphasize the need for very
accurate predictions for the theoretical power spectrum, pointing out
that currently used fitting functions such as the \citet{PD96} formula
or the fit derived by \citet{Smi03} underestimate the cosmic
shear-power spectra by $>30\%$ for $k>10\,h\,{\rm Mpc}^{-1}$.

\begin{figure}[t]
\centerline{
 \includegraphics[width=3in]{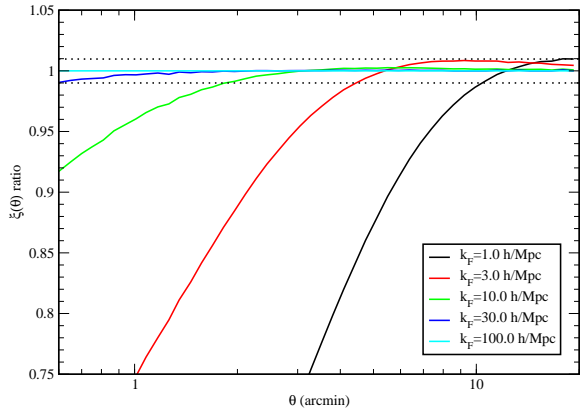}}
\caption{\label{theta} Ratio of the E-mode correlation function with
and without an assumed suppression of the power spectrum mimicking a
possible systematic error in the matter power spectrum. This figure
demonstrates that a gradual decrease in the accuracy of the matter
power spectrum on small scales will not lead to a catastrophic error
in the weak lensing prediction. The green line with
$k_F=10\,h\,{\rm Mpc}^{-1}$ corresponds to error properties which will be
close to the degradation we expect for the matter power spectrum
presented in this paper (see text).}
\end{figure}

We have independently assessed the impact a mis-modeled power spectrum
would have on the predictions of weak lensing observables, including
the fact that a wide range of spatial scales can be mapped into a
given angular scale. Assuming a distribution of sources with $\langle
z\rangle=1$, and using the Limber approximation, we compute the
observable shear-shear correlation function,
$\xi(\theta)=\langle\gamma(0)\cdot\gamma(\theta)\rangle$, given an
estimate of the $z$-dependent mass power spectrum, $\Delta^2(k,z)$. To
mimic the inaccuracy of $\Delta^2 (k,z)$ on scales smaller than
$1\,h\,{\rm Mpc}^{-1}$, we multiply it by a $z$-independent filter of
the form $(1+k^2/k_F^2)^{-1}$ for a variety of $k_F$. At $\ell=1000$
the suppression is 2-3\% for $k_F=1\,0\,h\,{\rm Mpc}$ ($k_F$ being the
assumed suppression of the power spectrum) and it drops to 1\% at
$\ell~500$. Assuming that $k_F\simeq 10\,h\,{\rm Mpc}^{-1}$ reflects
the error properties we are aiming at in this paper
(i.e.~$\Delta^2\sim 1\%$ low at $k\simeq\,1\,h\,$Mpc$^{-1}$ and
smoothly but increasingly low for smaller scales) we expect our
results could be used to predict the shear correlation function at the
percent level for separations larger than $2^\prime$.
Figure~\ref{theta} shows the expected error for different filter
scales. Assuming sources at higher $z$ shifts all of the curves to
larger scales, while a lower source redshift shifts the curves to
smaller scales.

In order to extract precise cosmological information from WL
measurements, additional physics beyond the gravitational contribution
must be taken into account. At length scales smaller than $k\sim
1\,h\,{\rm Mpc}^{-1}$, baryonic effects are expected to be larger than
1\% \citep{Whi04,ZhaKno04,Jin06,Rud08,GuiTeyCol09} and will have to be
treated separately, either directly via hydrodynamic simulations or,
as is more likely, by a combination of simulations and
self-calibration techniques (e.g.~constraining cluster profiles by
cluster-galaxy lensing at the same time as constraining the shear). In
any case, gravitational $N$-body simulations must remain the bedrock
on which all of these techniques are based.

Taking all of these considerations into account, the purpose of this
paper is to establish that gravitational $N$-body simulations can
produce power spectra accurate to 1\% out to $k\sim 1\,h\,{\rm
  Mpc}^{-1}$ between $z=0-1$ for a range of cosmological models. Given
the success of the CDM paradigm in explaining current observational
data we shall consider cosmologies within that framework. All of our
models will assume a spatially flat Universe with purely adiabatic
fluctuations and a power-law power spectrum. Since it is unlikely that
near-term observations can place meaningful constraints on the
temporal variation of the equation of state of the dark energy, we
will restrict attention to cosmologies with a constant equation of
state parameter $w=-p/\rho$ (where $p$ is the pressure and $\rho$ the
density of the dark energy with $w=-1$ in a $\Lambda$CDM cosmology).
Since $\Lambda$CDM is a good fit to the data, the accuracy of
simulations can be established primarily around this point.

In this paper we will establish that gravitational $N$-body
simulations can meet the above demands and derive a set of simulation
criteria which balance the need for accuracy against computational
costs. The target regime covers the most important range for current
and near-future WL surveys and additional physics is controllable at
the required level of accuracy. Showing that the required accuracy can
be obtained from $N$-body simulations is only the first step in
setting up a power spectrum determination scheme useful for weak
lensing surveys. In order to analyze observational data and infer
cosmological parameters, precise predictions for the power spectrum
over a large range of cosmologies are required. This paper --
establishing that achieving the base accuracy is possible -- is the
first in a series of three communications. In the second, we will
demonstrate that a relatively small number of numerically obtained
power spectra are sufficient to derive an accurate prediction scheme
-- or emulator -- for the power spectrum covering the full range of
desired cosmologies. The third paper of the series will present
results from the complete simulation suite, named the ``Coyote
Universe'' after the computing cluster on which it has been carried
out. The third paper will also contain a public release of a precision
power spectrum emulator.

In order to establish the accuracy over the required spatial dynamic
range, as well as over the redshifts probed, a variety of tests need
to be conducted. These include studies of the initial conditions,
convergence to linear theory at very large length scales, the mass
resolution requirement, and other evolution-specific requirements such
as force resolution and time-stepping errors. To establish robustness
of the final results, codes based on different $N$-body algorithms
should independently converge to the same results (within error
bounds). While some of these studies have been conducted separately
and within the confines of the cosmic code verification project
\cite[]{CodeCompare}, this is the first time that the more or less
complete set of possible problems has been investigated in realistic
simulations.

We find that it is indeed possible to control the accuracy of $N$-body
simulations at 1\% out to $k\sim 1\,h\,{\rm Mpc}^{-1}$. Even though
these scales are not very small, the simulation requirements are
rather demanding. First, the simulation volume needs to be large
enough to capture the linear regime accurately. Due to mode-mode
coupling, nonlinear effects influence scales as large as
500~$h^{-1}$Mpc. Therefore, the simulation volume needs to cover at
least 1~($h^{-1}$Gpc)$^3$. Second, with this requirement imposed, the
number of particles necessary to avoid errors from discreteness
effects at the smallest length scales of interest, also becomes
substantial. As we discuss later, because we are measuring the mass
power spectrum (which is sensitive to near-mean-density regions)
numerical results aiming for accuracy at the sub-percent level can
only be trusted at scales below the particle Nyquist wavenumber
\citep[see also][]{Joy08}. A $1\,(h^{-1}{\rm Gpc})^3$ simulation
volume requires a minimum particle loading of a billion particles.
Third, it is important to start the simulation at a high enough
redshift to allow enough dynamic range (in time) for structures to
evolve correctly and for the initial perturbations to be captured
accurately by the Zel'dovich approximation (ZA). Lastly, the force
resolution and time stepping has to be accurate enough to ensure
convergence of the simulation results.

The paper is organized as follows. In Section~\ref{chall} we use a
simple example to demonstrate the need for precision predictions from
theory. Section~\ref{nbody} contains a description of the $N$-body
codes used in this paper and some basic information about the
simulations. In Section~\ref{power} we briefly describe the power
spectrum estimator. In Sections~\ref{ics} and \ref{tests}
investigations of initial conditions and time evolution are reported,
demonstrating that the simulations can achieve the required accuracy
levels. Finally, we compare the numerical results to the commonly used
semi-analytic {\sc HaloFit\/} approach \citep{Smi03} in
Section~\ref{sec:halofit}, finding a discrepancy of $\sim 5-10\%$
between the fit and the simulations. We provide a summary discussion
of our results in Section~\ref{conclusion}. Appendix~\ref{app:2lpt}
discusses errors in setting up the initial conditions, comparing the
Zel'dovich and 2LPT approximations. Appendix~\ref{app:rich} provides
details of the Richardson extrapolation procedure used for some of the
convergence tests.

\section{The Precision Cosmology Challenge}
\label{chall}

Before discussing how to achieve 1\% accuracy for the nonlinear power
spectrum, we will briefly demonstrate the importance of accurately
determining the power spectrum. In our example, we assume the ability
to measure the power spectrum from observations at $1\%$ accuracy in
the quasi-linear and nonlinear regimes. On larger scales, accounting
for sample variance (statistical limitations due to finite
volume-sampling) leads to an increase in the statistical error, of up
to 10\%. These values are rough estimates, which are sufficient to
make our point in this simple example.

For our example, we use a halo model-inspired fitting formula given by
the code {\sc HaloFit\/} as implemented in
CAMB\footnote{http://camb.info}. Under the assumptions going into {\sc
  HaloFit\/} it can be straightforwardly modified for $w$CDM
cosmologies by simply adjusting the linear power spectrum and the
linear growth function to account for $w\not= -1$ (explicit tests for
some cosmologies were presented in \citealt{Ma07}). Current weak
lensing analyses (see, e.g., \citealt{kilb08}) rely on {\sc HaloFit\/}
to derive constraints for $w$CDM cosmologies due to the lack of a
better alternative. {\sc HaloFit\/} is therefore the natural choice
for our example.

\begin{figure}[t]
\centerline{
 \includegraphics[width=3in]{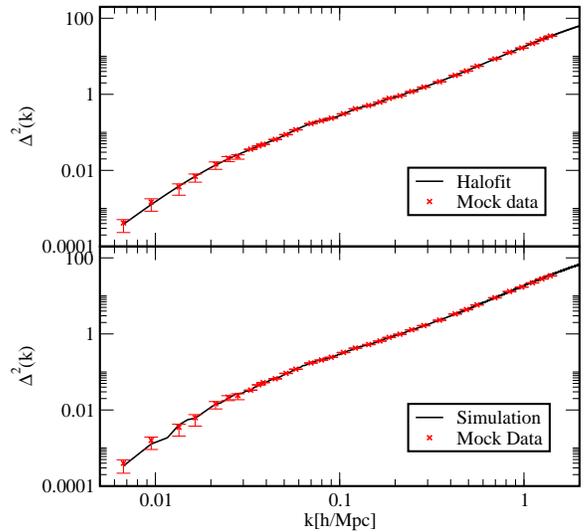}}
\caption{\label{mock}Upper panel: Synthetic data from a {\sc HaloFit\/}
  run. Lower panel: Synthetic data from a combination of several
  $N$-body runs. In both cases the black line shows the underlying
  power spectrum from which the data was drawn and the red points show
  34 data points with error bars. At small spatial scales, the assumed
  error is 1\%, rising to 10\% at large scales due to increased sample
  variance.}
\end{figure}

We generate two sets of mock measurements: one from a power spectrum
generated with {\sc HaloFit\/} and another directly from a set of
high-precision simulations. We then move points off the base power
spectrum according to a Gaussian distribution with variance specified
by the error estimates given above. The resulting mock data points and
the underlying power spectra are shown in Figure~\ref{mock}. On a
logarithmic scale, the data points and power spectra are almost
indistinguishable. As we will show later in Section~\ref{sec:halofit},
the difference between the {\sc HaloFit\/} and $N$-body power spectra
is at the 5-10\% level: this difference is enough to lead to
significant biases in parameter estimation.

\begin{figure}[t]
\centerline{
 \includegraphics[width=3in]{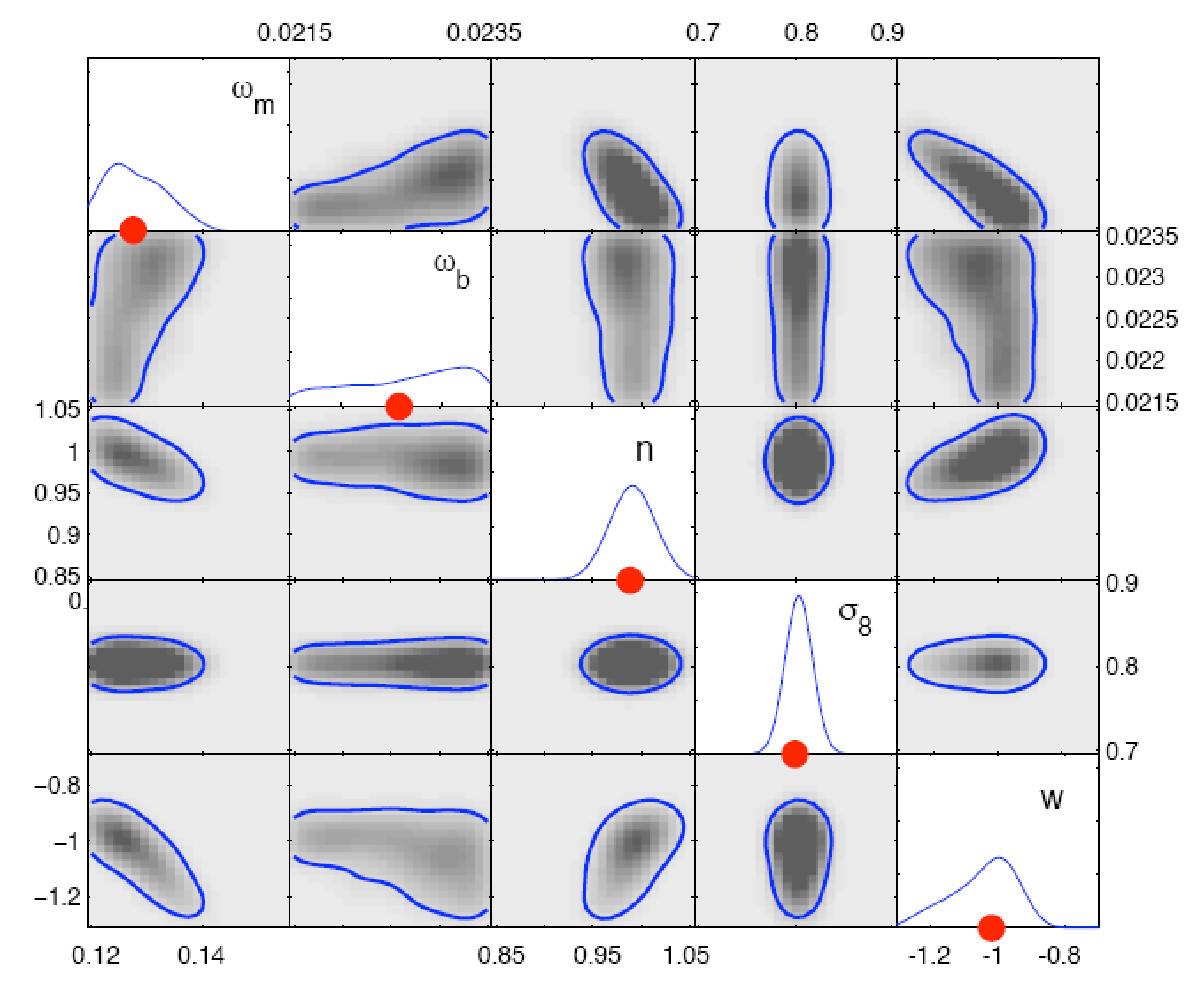}}
\vspace{0.2cm}
\centerline{
 \includegraphics[width=3in]{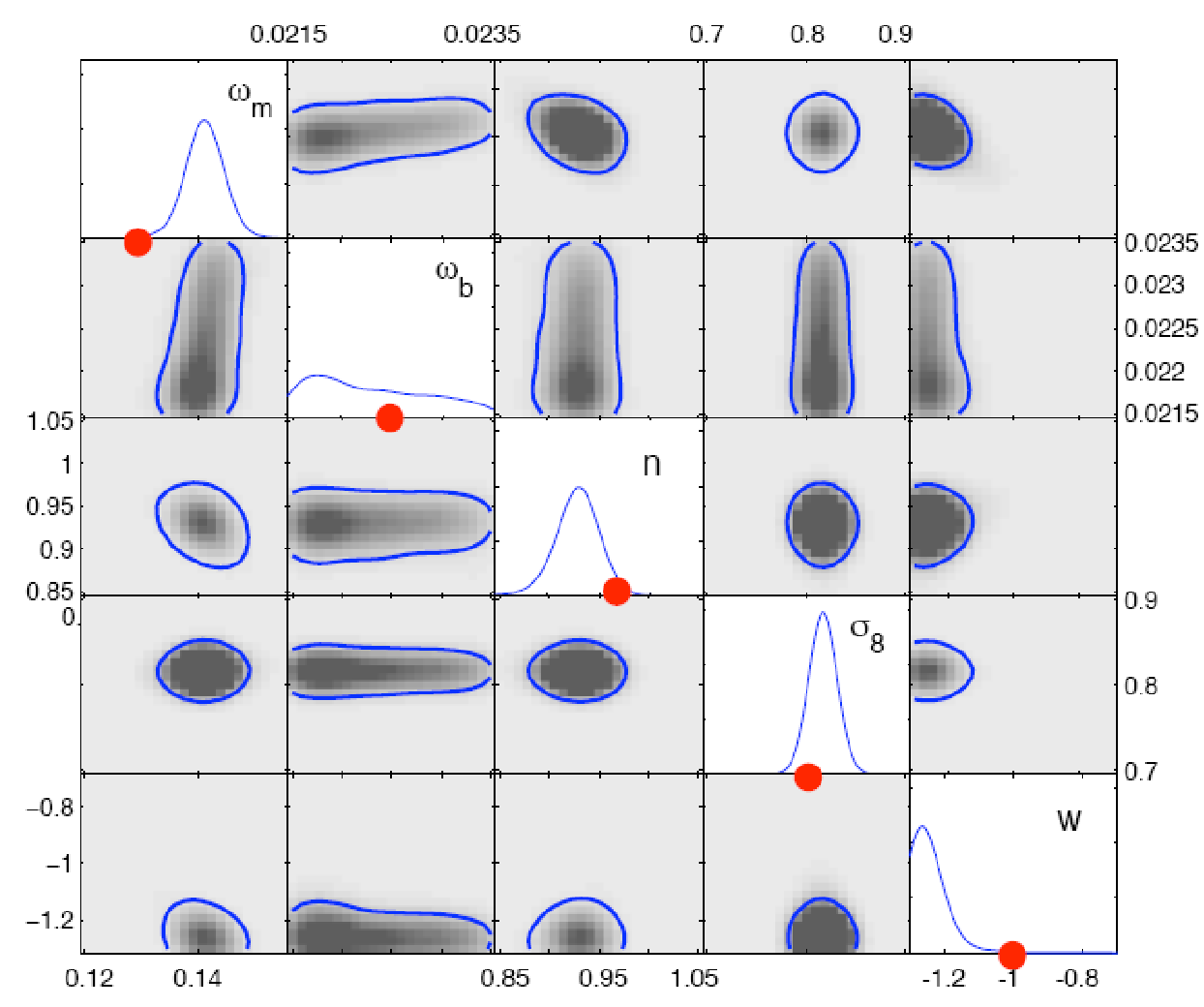}}
\caption{\label{likelihood}Posterior distributions for the five
  parameters under consideration. Upper panel: Results for the
  analysis of the {\sc HaloFit\/} synthetic data set analyzed with a set
  of {\sc HaloFit\/} power spectra. The red dots indicate the true
  values. As is to be expected, the constraints on the parameters are
  very good. Lower panel: Results for the {\sc HaloFit}-based analysis
  of the $N$-body synthetic data set. Note that the constraints for
  $\omega_m$ and $w$ are now incorrect at  $\sim 20$\%.}
\end{figure}

We determine the best-fit parameters from the two mock data sets using
the following parameter priors:
\begin{eqnarray}
0.02\le &\omega_b&\le 0.025,\nonumber\\
0.11\le &\omega_m&\le 0.15,\nonumber\\
0.85\le &n_s&\le 1.05,\nonumber\\
-1.3\le &w&\le -0.7,\nonumber\\
0.7\le &\sigma_8&\le 0.9,
\label{priors}
\end{eqnarray}
where $\omega_b=\Omega_b h^2$ and $\omega_m=\Omega_m h^2$. We do not
treat $h$ as an  independent variable but determine it via the CMB
constraint $l_A=\pi d_{lss}/r_s=302.4$ where $d_{lss}$ is the distance
to the last scattering surface and $r_s$ is the sound horizon (more
details of how we construct our model sampling space are provided in
\citealt{Hei09}, Paper II).

The parameter estimation analysis then proceeds via a combination of
model interpolation and Markov Chain Monte Carlo (MCMC) as implemented
in our recently introduced cosmic calibration framework \citep{Hab07}.
We use {\sc HaloFit\/} to generate the nonlinear power spectra for the
MCMC analysis. That is, we analyze a {\sc HaloFit\/} synthetic data
set and one generated from numerical simulations against a set of
model predictions from {\sc HaloFit\/} generated power spectra. The
results, which are all obtained from data at $z=0$, are shown in
Figure~\ref{likelihood}. The upper panel shows the results from the
analysis of the {\sc HaloFit\/} synthetic data, where the parameter
estimation works extremely well, being essentially a consistency check
for the statistical framework. The result also points to the
constraining power of matter power spectrum data. The lower panel in
Figure~\ref{likelihood} shows the corresponding result for the
synthetic data generated directly from the simulations. In this case,
the $\sim 5\%$ errors in the {\sc HaloFit\/} model predictions are
clearly seen to be problematic: most of the parameters are
significantly off, $\omega_m$ and $w$ being mis-estimated by $\sim
20$\%.

The example used here is certainly too simplified, relying only on
large scale structure ``observations'' and making no attempt to take
into account covariance, degeneracies, other observations, etc. For
example, including a second observational probe such as the cosmic
microwave background would provide a tighter constraint on $\sigma_8$,
reducing the 20\% shift in $w$. Nevertheless, the example clearly
illustrates the general point that to perform an unbiased data
analysis the theory underlying the analysis framework must match or
preferably exceed the accuracy of the data.

\section{$N$-body codes and Simulations}
\label{nbody}

\begin{figure}[t]
\begin{center}
\resizebox{3in}{!}{\includegraphics{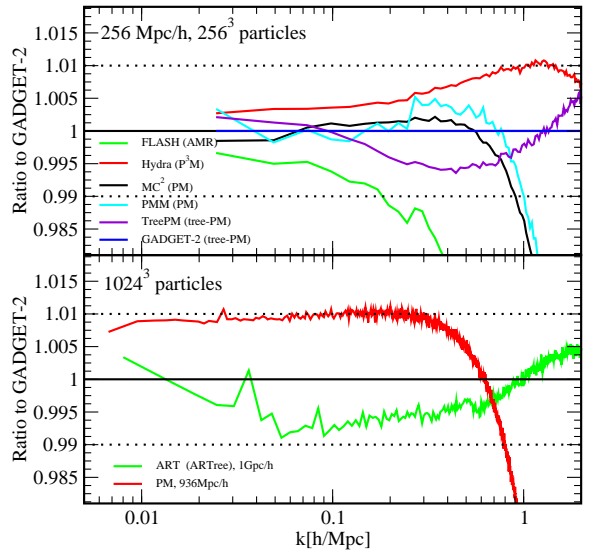}}
\end{center}
\caption{Upper panel: Comparison of dimensionless power spectra from a
  handful of $N$-body codes, taken from the data of
  \protect\citet{CodeCompare} for the ``LCDMb'' box: a $\Lambda$CDM
  model with $\Omega_m=0.314$, $h=0.71$, $n_s=0.99$, and $L_{\rm
    box}=256\,h^{-1}$Mpc with 256$^3$ particles. The two PM codes,
  MC$^2$ and PMM, were run on a 1024$^3$ grid (with a grid-to-particle
  ratio of 4:1, a factor of two higher than used for the PM runs in
  this paper). The FLASH run had a base grid of 256$^3$ and a
  refinement level of two. Therefore, the force resolution of the
  purely grid-based codes is roughly a factor of ten lower than 
  for the other codes (the different force kernels make a precise
  comparison difficult). The dotted lines show the 1\% agreement
  limit. The high force-resolution codes agree to ${\mathcal O}(1\%)$
  up to $k\sim 1\,h\,{\rm Mpc}^{-1}$ despite different choices for the
  force softening and other numerical parameters. Lower panel:
  Comparison of  {\small GADGET-2\/} and ART for a simulation with
  1024$^3$ particles and $L_{\rm box}=1~h^{-1}$Gpc. The cosmological
  parameters are very close to those for our major runs, the main
  difference being the starting redshift of $z_{\rm in}=65.66$. The
  agreement of the two codes is better than 1\% over all scales. In
  addition, we compare one of the PM runs used in this paper with
  respect to {\small GADGET-2}. The agreement is also at ${\mathcal
    O}(1\%)$. We re-emphasize that our goal is to derive simulation
  requirements for  percent level accuracy and finding a good balance
  between efficient computing and accuracy. By tuning code parameters,
  the agreement between different codes may be improved, but this
  would defeat the purpose of testing for robustness.
\label{fig:codecompare}}
\end{figure}

The numerical computations carried out and analyzed in this paper are
$N$-body simulations that model structure formation in an expanding
universe assuming that gravity dominates all other forces. The phase
space density field is sampled by finite-mass particles and these
particles are evolved using self-consistent force evaluations.
Although the effects of baryons and neutrinos are taken into account
while setting up initial conditions, only their gravitational
contribution to the ensuing nonlinear dynamics of structure formation
is kept (along with that of the dark matter). Gas dynamics, feedback
effects, etc.~are all neglected. At sufficiently small scales this
neglect is clearly not justified, but at the 1\% level and for
wavenumbers smaller than $k\sim 1\,h\,{\rm Mpc}^{-1}$ this assumption
is expected to hold.

In order to solve the $N$-body problem, we employ two commonly used
algorithms, the particle-mesh (PM) approach and the tree-PM approach.
The $N$-body methods model many-body evolution problems by solving the
equations of motion of a set of tracer particles which represent a
sampling of the system phase space distribution. In PM codes, a
computational grid is used to increase the efficiency of the
self-consistent inter-particle force calculation. In the codes used in
this paper, the Vlasov-Poisson system of equations for an expanding
universe is solved using Cloud-in-Cell (CIC) mass deposition and
interpolation with second order (global) symplectic time-stepping and
a Fast Fourier Transform (FFT)-based Poisson solver. The advantage of
the PM method is good error control and speed, the major disadvantage
is the restriction on force resolution imposed by the biggest FFT that
can be performed (typical current limits being 2048$^3$ grids or
4096$^3$ grids). Two independently written PM codes were checked
against each other in the low $k$ regime, one being the PM code MC$^2$
described in \citet{Hei05}, with excellent agreement being achieved.
In addition, the publicly available code {\small GADGET-2\/}
\citep{Spr05} was slightly modified to run in pure PM mode. The
agreement between these codes was excellent.

Tree-PM is a hybrid algorithm that combines a long-range force
computation using a grid-based technique, with shorter-range force
computation handled by a tree algorithm. The tree algorithm is based
on the idea that the gravitational potential of a far-away group of
particles is accurately given by a low-order multipole expansion.
Particles are first arranged in a hierarchical system of groups in a
tree structure. Computing the potential at a point turns into a
descent through the tree. For most of our high-resolution runs we use
the tree-PM code {\small GADGET-2\/}, for some of the tests and
comparison we also use the code TreePM which is described in
\citet{Whi02}.

Several different $N$-body codes have been compared in previous
work~\cite[]{Hei05,CodeCompare}, including PM, tree-PM,
adaptive-mesh-refinement, pure tree, and particle-particle PM codes.
The results of these code verification tests are consistent with the
idea that 1\% error control is possible up to $k\sim 1\,h\,{\rm
  Mpc}^{-1}$ (at $z=0$), as shown in Figure~\ref{fig:codecompare}. The
upper panel in the figure shows a comparison of the power spectra from
a subset of the codes used in \citet{CodeCompare} with respect to a
{\small GADGET-2\/} run. The simulations are performed with 256$^3$
particles in a 256~$h^{-1}$Mpc box. We find agreement at the
one-percent level between the high resolution codes despite the use of
different choices for the force softening and other numerical
parameters. In a separate test, we compared {\small GADGET-2} with the
Adaptive Refinement Tree (ART) code \citep{Kra97,GotKly08}. The
simulation encompassed a volume of (1~$h^{-1}$Gpc)$^{3}$ and 1024$^3$
particles. The agreement between the two codes was again better than
one percent between $z=0$ and $z=1$ and out to $k\sim 1\,h\,{\rm
  Mpc}^{-1}$. The result for $z=0$ is shown in the lower panel of
Figure~\ref{fig:codecompare}. The excellent and robust -- w.r.t.
numerical parameter choices -- agreement between different codes
provides confidence that it is possible to predict the matter power
spectrum at the desired accuracy.

We use a combination of PM and tree-PM runs for this paper, and in the
follow-up work, to create an accurate prediction for the matter power
spectrum. At quasilinear spatial scales -- large, yet not fully
described by linear theory ($k\sim0.1\,h\,{\rm Mpc}^{-1}$) -- lower
resolution PM simulations are adequate. Furthermore, to reduce the
variance due to finite volume-sampling -- a problem at low values of
$k$ -- simulations should be run with many realizations of the same
cosmology. We fulfill this requirement by running a large number of PM
simulations with either $512^3$ or $1024^3$ particles. In order to
resolve the high-$k$ part of the power spectrum, we use the {\small
  GADGET-2} code.

The codes are run with different settings as explicitly discussed in
the tests mentioned below. In the case of the {\small GADGET-2} runs,
we use a PM grid twice as large, in each dimension, as the number of
particles, and a (Gaussian) smoothing of $1.5$ grid cells. The force
matching is set to $6$ times the smoothing scale, the tree opening
criterion being set to $0.5\%$. The softening length is set to 50~kpc.
For more general details on the code settings in {\small GADGET-2\/}
and the code itself, see \citet{Spr05}.

The pure PM simulations have twice as many mesh points in each
dimension as there are particles. The integration variables are the
position and conjugate momentum, with time-stepping being in constant
steps of $\Delta\ln a=0.02$. The forces are obtained using $4^{\rm
  th}$ order differencing from a potential field computed using
Fourier transforms. The input density field is obtained from the
particle distribution using CIC charge assignment \citep{HocEas89} and
the potential is computed using a $1/k^2$ kernel.

If not stated otherwise, our fiducial $\Lambda$CDM model has the
following cosmological parameters: $\Omega_m=0.25$ for the total
matter content, a cosmological constant contribution specified by
$\Omega_\Lambda =0.75$, baryon density as set by $\omega_b=\Omega_b
h^2=0.024$, a dimensionless Hubble constant of $h=0.72$, the
normalization specified by $\sigma_8=0.8$, and a fixed spectral index,
$n_s=0.97$. These parameters are in accord with the latest WMAP
results~\cite[]{WMAP5}. The model is run with box size of
(936~$h^{-1}$Mpc)$^3$ and with 1024$^3$ particles. For some of the
tests we use a downscaled version of this simulation but keep the
inter-particle spacing approximately the same (1$\,h^{-1}$Mpc).

\section{Power spectrum estimation} \label{power}

The key statistical observable in this paper is the density
fluctuation power spectrum $P(k)$, the Fourier transform of the
two-point density correlation function. In dimensionless form, the
power spectrum may be written as
\begin{equation}
  \Delta^2(k) \equiv \frac{k^3P(k)}{2\pi^2},
\label{delta}
\end{equation}
which is the contribution to the variance of the density
perturbations per $\ln k$.  

Because $N$-body simulations use particles, one does not directly
compute $P(k)$ or equivalently, $\Delta^2(k)$. Our procedure is to
first define a density field on a grid with a fine enough resolution
such that the grid filtering scale is much higher than the $k$ scale
of interest. This particle deposition step is carried out using CIC
assignment. The application of a discrete Fourier transform (FFT) then
yields $\delta(\bf{k})$ from which we can compute
$P(\bf{k})=|\delta(\bf{k})|^2$, which in turn can be binned in
amplitudes to finally obtain $P(k)$. Since the CIC assignment scheme
is in effect a spatial filter, the smoothing can be compensated by
dividing $P(\bf{k})$ by $W^2(\bf{k})$, where
\begin{equation}
  W\left(\bf{k}\right) =
    j_0^2\left(\frac{k_xL_g}{2}\right)
    j_0^2\left(\frac{k_yL_g}{2}\right)
    j_0^2\left(\frac{k_yL_g}{2}\right),
\end{equation}
and $L_g$ is the size of the grid cell. Typically the effect of this
correction is only felt close to the maximum (Nyquist) wavenumber for
the corresponding choice of grid size. One should also keep in mind
that particle noise and aliasing artifacts can arise due to the finite
number of particles used in $N$-body simulations and due to the finite
grid size which is used for the power spectrum estimation. As
explained further below, convergence tests based on varying the number
of sampling particles can help establish the smallest length scales at
which accurate results can be obtained. The particle loading in our
simulations is sufficient to resolve the power spectrum at the scales
of interest, such that possible shot noise is at the sub-percent
level.

It is common to make a correction for finite particle number by
subtracting a Poisson ``shot-noise'' component from the bin-corrected
power spectrum:
\begin{equation}\label{shot}
   \Delta^2_{\rm shot}(k) = \frac{k^3}{2\pi^2}\left(\frac{L}{N_p}\right)^3,
\end{equation}
where $N_p$ is the cube-root of the number of particles and $L$ is the
box length.  We have not done this in this paper because our particle
loading is large enough to render it a small correction on the scales
of interest and it is not clear that this form captures the nature of
the correction correctly.  Note that the initial conditions have
essentially no shot noise at all, and the evolution prior to
shell-crossing does not add any.  Shot noise thus enters through the
high-$k$ sector and filters back to lower $k$ in a complex manner.

\begin{figure*}[t]
\begin{center}
\resizebox{5.5in}{!}{\includegraphics[angle=270]{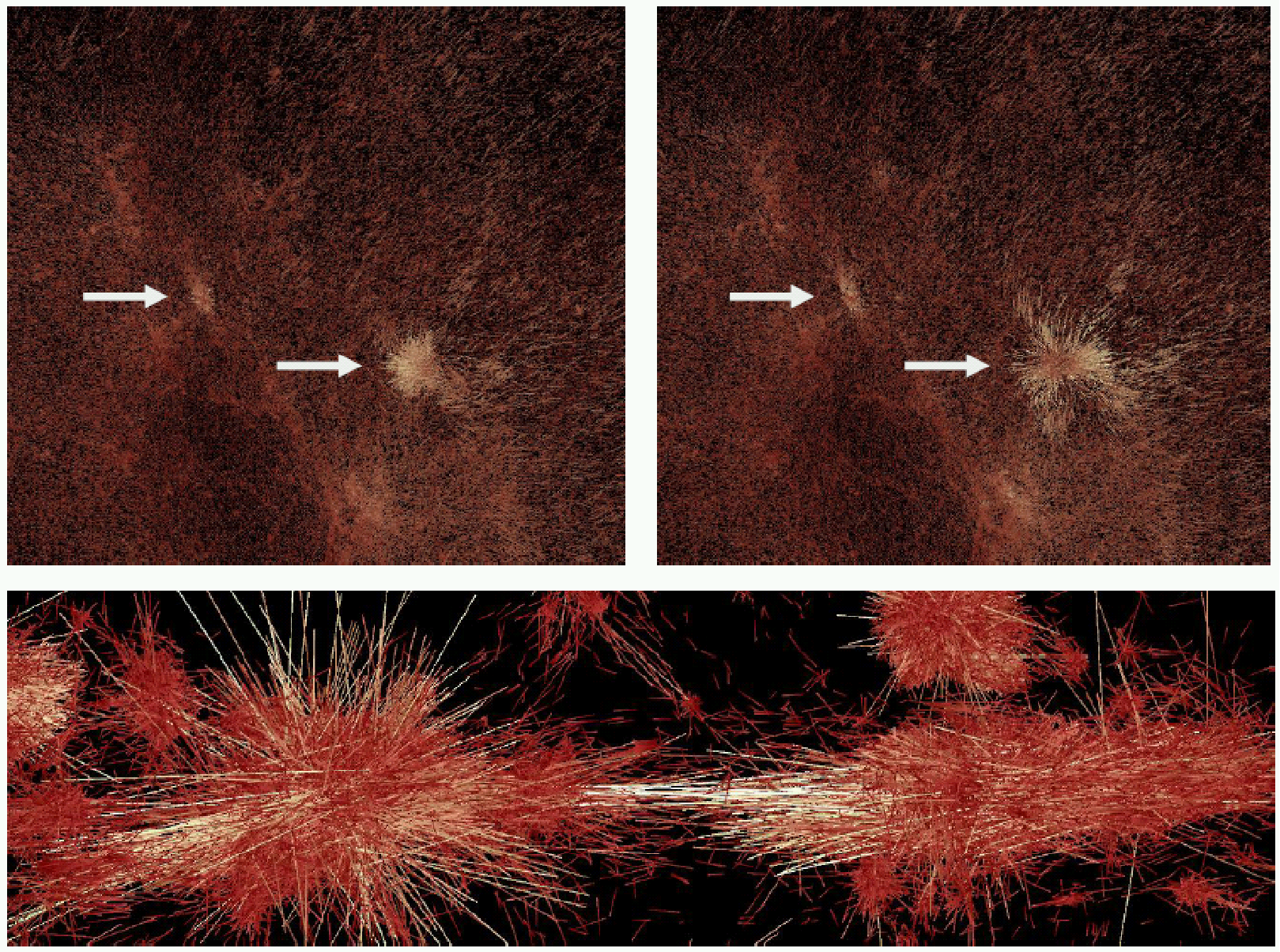}}
\end{center}
\caption{Top two panels: comparison of outputs at $z=10$ using
  different starting redshifts. The particles are colored with respect
  to their velocities. The simulation box is 8$\,h^{-1}$Mpc on a side. The
  simulation shown in the left panel was started at $z_{\rm in}=250$,
  while for the other, $z_{\rm in}=50$. In the simulation started at
  $z_{\rm in}=50$, structures formed by $z=10$ are not as concentrated
  as in simulations with a high-$z$ start, leading to the possible
  lowering of halo masses. The lower panel shows differences along a
  filament. In this case a line was drawn between each particle
  position in the two different data sets. The longer  the line, the
  larger the difference due to the two different initial redshifts.
  For more details see \protect\citet{Har08,HarHei08,Luk07}.} 
\label{fig:vis}
\end{figure*}

We average $P(\bf{k})$ in bins linearly spaced in $k$ of width $\Delta
k\simeq 0.001\,{\rm Mpc}^{-1}$, and report this average for each bin
containing at least one grid point. We assign to each bin the $k$
associated with the unweighted average of the $k$'s for each grid
point in the bin. Note that this procedure introduces a bias in
principle, since for nonlinear functions $\langle f(x)\rangle\ne
f(\langle x \rangle)$, but our bins are small enough to render this
bias negligible.

In a recent paper, \citet{Col08} suggest an alternative approach to accurately
estimate power spectra from $N$-body simulations. Their method is
based on a Taylor expansion of trigonometric functions as a
replacement for large FFTs. The idea is to estimate the power spectrum out
to small scales with minimal memory overhead, a major obstacle for the
brute force FFT approach. We have checked their method up to fifth
order against our results from the $2048^3$ FFT and found excellent
agreement. Our FFT is clearly large enough to avoid any aliasing at
$k\sim 1\,h\,{\rm Mpc}^{-1}$. 

\section{Initial conditions}
\label{ics}

The initial conditions in $N$-body codes are often a source of
systematic error in ways that can sometimes be hard to detect. It is,
therefore, essential to ensure that the implementation of the initial
conditions is not a limiting factor in attaining the required accuracy
of the power spectrum over the redshift range of interest. An
important aspect here is the choice of starting redshift. There are
two reasons for this: (i) The Lagrangian perturbation theory used to
generate the initial particle distribution (usually the leading order
Zel'dovich approximation) is more accurate at higher redshifts, and
(ii) for a given (nonlinear) $k$ scale of interest, enough time must
have elapsed for the correct nonlinear power spectrum to be
established at that scale, at the redshift of interest.

Due to a combination of the two effects mentioned above, delayed
starts typically lead to a suppression of structure formation
(including the halo mass function) as shown in Figure~\ref{fig:vis}.
We now describe our basic methodology for generating initial
conditions and choosing the starting redshift.

\subsection{Initial Condition Generation}
\label{icgen}

As is standard, we generate our initial conditions by displacing
particles from a regular Cartesian grid (``quiet start'') using the
Zel'dovich approximation~\cite[]{Zel70}. In this approximation, the
particle displacement and velocity are given by
\begin{eqnarray}
\label{zeldo}
{\bf x}(\bf{q})&=&{\bf{q}}-D_1{{\bf\nabla}}_q\phi^{(1)},\\
{\bf v}=\frac{d{\bf x}}{dt}&=&-D_1f_1H{{\bf\nabla}}_q\phi^{(1)}.
\end{eqnarray}
Here $\bf{q}$ is the initial (on-grid) position of the particle,
$\bf{x}$ is the final position, $D_1$ is the linear growth factor
defined below in Eqn.~(\ref{dplus}) and $\phi^{(1)}$ is the potential
field. $H$ is the Hubble constant, $f_i$ is the logarithmic derivative
of the growth function $f_i=(d\ln D_i)/(d\ln a)$, and the
time-independent potential $\phi^{(1)}$ obeys the Poisson equation
${\bf \nabla}^2_q\phi^{(1)}({\bf{q}})=\delta({\bf{q}})$.

A recent suggestion is to determine the initial displacement of the
particles and their velocities via second order Lagrangian
perturbation theory (2LPT) instead of using the (leading order)
Zel'dovich approximation \citep{Sco98,Cro06}. In principle, this could
allow a later start of the simulation (lower $z_{\rm in}$) without
losing accuracy in the final result. However, it does not address the
problem of keeping a sufficient number of expansion factors between
the initial and final redshifts. Additionally, error control of the
perturbation theory and its convergence properties need to be
carefully checked. We have therefore decided on a more conservative
approach: instead of using higher order schemes to generate initial
conditions, we choose a high enough starting redshift that higher
order effects are negligible (see Appendix \ref{app:2lpt}). Since most
of the code's runtime is at low redshift, the additional overhead for
starting the simulation early is minimal.


The potential field is generated from a realization of a Gaussian random
density field $\delta(\bf{k})$ (with random phases). The initial power
spectrum is 
\begin{equation}
  P(k)=Bk^nT^2(k),
\end{equation}
where $B$ determines the normalization and $T(k)$ is the matter transfer
function. We compute $T(k)$ using the numerical code CAMB. The results
from CAMB were compared against those generated by an independent code
described in \citet{WhiSco96}, \citet{HuWhi97} and \citet{HSWZ}. The
results from this code are known to agree well with
CMBfast~\cite[]{CMBcompare}. The final level of agreement was at the
$\sim 10^{-3}$ level for the $k$ modes of interest, comfortably below
our 1\% goal. 

The displacement field is easily generated in Fourier space: The
Fourier transform of the displacement field is proportional to
$({\bf k}/k^2)\delta({\bf k})$ in the continuum, and we compute the
displacements using FFTs. The FFT grid is chosen to have twice as many
points, in each dimension, as there are particles.

The scale-independent linear growth factor, $D_1(z)$, satisfies
\citep[e.g.][]{PeacockBook}
\begin{equation}
\label{dplus}
  \left(\frac{D_1}{a}\right)'' +
  \left( 4+\frac{1}{2}\frac{\rho_c'}{\rho_c}\right)
  \left(\frac{D_1}{a}\right)' - 
  \left( \frac{3}{2}\frac{\rho_m}{\rho_c}-
  \frac{1}{2}\frac{\rho_c'}{\rho_c}-3\right)
\left(\frac{D_1}{a}\right) = 0. 
\end{equation}
Here $\rho_c\propto H^2$ is the critical density, $\rho_m$ is the
matter density and primes denote differentiation with respect to $\ln
a$.  Our convention has $D_1(z=0)\equiv 1$ and $D_1(z)\propto
(1+z)^{-1}$ when $\rho_m\simeq\rho_c$.  This procedure neglects the
differential evolution of the baryons and dark matter, but since we
are simulating only collisionless systems here this is the most
appropriate choice. 
Future simulations including baryons will have to deal with this
question in more detail.

\subsection{The Initial Redshift}

The choice of the starting redshift depends on three factors: the
simulation box size, the particle loading, and the first redshift at
which results are desired. The smaller the box and the higher the
first redshift of interest, the higher the initial redshift must be.
It is not easy to provide a universal ``recipe'' for determining the
optimal starting redshift. For each simulation set-up, convergence
tests must be performed for the quantities of interest. Nevertheless,
there are several guiding principles to determine the starting
redshift for a given problem. These are:

\begin{itemize}
\item Ensure that any unphysical transients from the initial
  conditions are negligible at the redshift of interest.
\item  Ensure a sufficient number of expansion factors to allow
  structures to form correctly at the scales of interest. 
\item Ensure that the initial particle move on average is much smaller
  than the initial inter-particle spacing.
\item Ensure that $\Delta^2(k)\ll 1$ at the wavenumber of interest.
\end{itemize}
 
A more detailed description -- from a mass function-centric point of
view -- can be found in \citet{Luk07}. The aim here is to measure the
power spectrum from a (936~$h^{-1}$Mpc)$^3$ box between $z=1$ and
$z=0$ at $k=1\,h\,{\rm Mpc}^{-1}$ at 1\% level accuracy. In order to
fulfill the first and second criteria given above, we generate the
initial conditions at $z_{\rm in}$ such that $D(z_{\rm
  in})/D(z=1)=0.01$. With $D_1(z)\simeq a(z)=1/(1+z)$ this leads to a
starting redshift of approximately $z_{\rm in}=200$ and one hundred
expansion factors between the starting redshift and $z=1$. Note that
this criterion is completely independent of the box size and particle
loading, though it is cosmology dependent via the growth rate.

For the ($936\,h^{-1}$Mpc)$^3$ boxes we simulate, this starting
redshift leads to rms displacements between $3-5\%$ of the mean
inter-particle spacing, satisfying the condition that the rms
displacement should be much less than the mean inter-particle spacing.
This measurement clearly depends on the box size. A smaller box would
have led to much bigger displacements with respect to the mean
inter-particle spacing. At $z_{\rm in}=200$ the dimensionless power at
the fundamental mode is $\mathcal{O}(10^{-8})$ and at the Nyquist
frequency is $\mathcal{O}(10^{-4})$ which clearly satisfies the last
point of the list above. We show a series of convergence tests
including a higher order Lagrangian scheme in Appendix~\ref{app:2lpt}.

\section{Resolution Tests}
\label{tests}

In order to ensure that our results are properly converged for $k\le
1\,h\,{\rm Mpc}^{-1}$ between $z=1$ and $z=0$ we need to understand
the impact of box size, particle loading, force softening, and
particle sampling on the numerically determined power spectra.

\subsection{Box Size}

\begin{figure}
\centerline{
 \includegraphics[width=3.4in]{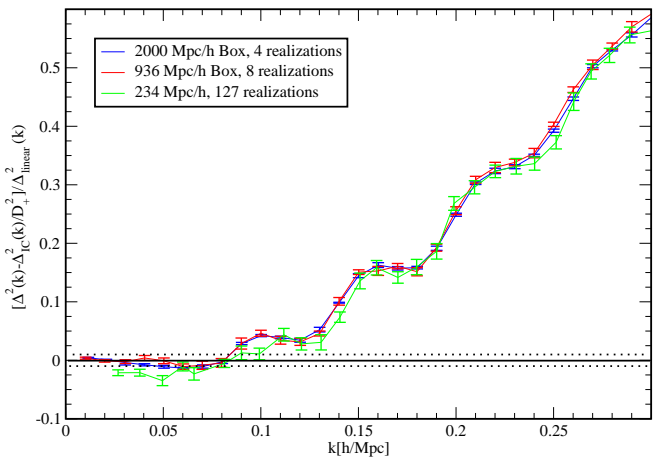}}
\caption{\label{box}
Comparison of power spectra from three different box sizes,
$2000\,h^{-1}$Mpc (blue) and $936\,h^{-1}$Mpc (red), and
$234\,h^{-1}$Mpc (green). We subtract the initial power spectrum scaled
to $z=0$ from the final power spectrum, average over all realizations
(4 for the big box, 8 for the medium box, and 127 for the small box),
and divide the result by the smooth prediction from linear theory. The
error bars show the standard deviation. The overall agreement of
$\Delta^2(k)$ from the two large box sizes is better than 1\% on
scales below $k\sim0.1\,h\,{\rm Mpc}^{-1}$ (the 1\% limit is shown by
the dotted lines). The small box result displays an overall
suppression of the power spectrum at low $k$ (see text).}
\end{figure}

The choice of the box size depends on several factors. In principle,
one should choose as large a volume as practicable, to ensure that the
largest-scale modes are (accurately) linear at the redshift of
interest (in our case between $z=1$ and $z=0$), improve the
statistical sampling (especially for BAO), and to obtain accurate
tidal forces. If the box volume is too small, the largest modes in the
box may still appear linear at the redshift of interest, even though
they should have already gone nonlinear. This leads to a delayed onset
of the nonlinear turnover and the quasi-linear regime is treated
incorrectly.

Practical considerations, however, add two restrictions to the box
size arising from (i) the necessarily finite number of particles, and
for the PM simulations, (ii) limitations on the force resolution. The
storage requirements and run time for the N-body codes scale (close
to) linearly with particle number, so running many smaller boxes
``costs'' as much as running one very large box with more particles.
However the ability to move jobs through the queue efficiently and
post-process the data all argue in favor of more smaller jobs than one
very large job.

The CDM power spectrum peaks roughly at $k\sim0.01\,h\,{\rm
  Mpc}^{-1}$, determined by the horizon scale at the epoch of
matter-radiation equality. As the power falls relatively steeply below
this value of $k$, a box size of $1\,(h^{-1}$Gpc)$^3$, corresponding
to a fundamental mode of $k\sim0.006\,h\,{\rm Mpc}^{-1}$, is a
reasonable candidate for comparing with linear theory on the largest
scales probed in the box. [These considerations are of course redshift
and $\sigma_8$-dependent: at $z=0$, small nonlinear mode-coupling
effects can be seen below $k\sim0.1\,h\,{\rm Mpc}^{-1}$ (cf.~Figure
\ref{box}). At higher redshifts, these effects move to higher $k$.] Of
course, bigger boxes are even better (especially for improved
statistics, although this is unrelated to linear theory
considerations), and a convergence test in box size is described
below.

The particle loading is particularly significant as it sets the
maximum wavenumber below which the power spectrum can be accurately
determined. As discussed in Section~\ref{massr}, the accuracy of the
power spectrum degrades strongly beyond the Nyquist wavenumber, which
depends on both the box size and particle number [see
Eqn.~(\ref{nyquist})]. Therefore, a compromise has to be found between
box size and particle loading. After having decided the size of the
smallest scale of interest and the maximum number of particles that
can be run, the box size is basically fixed. In our case, the optimal
solution (considering computational resources) appears to be a box
size of roughly 1~$h^{-1}$Gpc on a side and a particle loading of one
billion particles -- covering a wavenumber range $0.0067\,h\,{\rm
  Mpc}^{-1}<k<3.4\,h\,{\rm Mpc}^{-1}$ with the upper limit given by
the Nyquist wavenumber.

The force resolution for PM codes is a direct function of the box
size, once the size of the density (or PM) grid is fixed. While other
codes do not have this restriction in principle, PM codes are very
fast, and have predictable error properties. In order to obtain
sufficient statistics and accuracy for determining $P(k)$, results
from many large volume runs at modest resolution can be ``glued'' to
those from fewer high resolution runs, providing an optimal way to
sample the quasilinear and nonlinear regimes. PM simulations are very
well suited to handling the quasilinear regime; for a Gpc$^3$ box, a
2048$^3$ grid provides enough resolution to match the high-resolution
runs out to $k\sim0.5\,h\,{\rm Mpc}^{-1}$.

In order to ensure that a Gpc$^3$ box is sufficient to obtain accurate
results on very large scales, we compare the results from 8
realizations in a (936~$h^{-1}$Mpc)$^3$ box, 4 realizations in a
(2000~$h^{-1}$Mpc)$^3$ box, and 127 realizations in a
(234~$h^{-1}$Mpc)$^3$ box. The large volume runs were run with
1024$^3$ particles on a 2048$^3$ grid each, the smaller volumes were
run with 512$^3$ particles on a 1024$^3$ grid. We subtract the power
spectrum from the initial redshift scaled by the growth factor to
$z=0$ from the final power spectrum, average over all realizations and
divide by the linear theory answer. The results are shown in
Figure~\ref{box}. The agreement between the two sets of large volume
simulations is much better than one percent. The agreement with linear
theory on scales below $k\sim0.1\,h\,{\rm Mpc}^{-1}$ is roughly at the
percent level and much better than this for $k\sim0.01\,h\,{\rm
  Mpc}^{-1}$. We note that for the cosmology used in our study, we do
not observe a suppression of the power spectrum with respect to linear
theory by $\sim 5\%$ on scales of $0.05\,h\,{\rm Mpc}^{-1} < k <
0.075\,h\,{\rm Mpc}^{-1}$ as was reported in, e.g., \citet{Smi07}. The
results for the smaller boxes is a few percent below linear theory at
large scales and the onset of the nonlinear regime is captured
inaccurately. Thus, small box simulations suffer from two defects:
first, a large number of simulations is required to overcome finite
sampling scatter at low $k$, and, second, all simulations are biased
low due to the unphysical suppression of the power spectrum amplitude.

In a recent paper, \citet{Tak08} discuss finite volume effects in
detail and propose a way to use perturbation theory to eliminate these
effects. They have two concerns: (i) A small simulation volume will
lead to enhanced statistical scatter on large scales, if only a few
realizations are considered. (ii) If the simulation volume is too
small and the linear regime is not captured accurately, the result for
the power spectrum will be biased low. We overcome the first
difficulty by running many realizations of our cosmological model. In
combination with our large simulation volume, we are able to keep the
statistical noise below the percent level. The second concern is
clearly valid if the simulation box is too small. With the Gpc$^3$ and
larger volumes we consider, no size-related bias is observed. The two
different box sizes we investigate are in good agreement as can be
seen in Figure~\ref{box}. One concern with respect to the
\citet{Tak08} results is that they start their simulations rather late
($z_{\rm in}=30$) and investigate the results starting at $z=3$. As
demonstrated in Figure~\ref{fig:ic_z0} such a late start suppresses
the power spectrum at quasilinear and nonlinear scales.

\subsection{Mass Resolution}
\label{massr}

\begin{figure}[t]
\centerline{
 \includegraphics[width=3in]{sampling.eps}}
\caption{\label{fig1}Importance of particle sampling for calculating
  the power spectrum. The underlying simulation is the {\small
  GADGET-2} run with 1024$^3$ particles. All power spectra are
  measured on a 2048$^3$ grid. Upper panel: green -- power spectrum
  from 1024$^3$ particles at $z=0$, red -- from the 512$^3$
  downsampled distribution, blue -- from the 256$^3$ downsampled
  distribution, black -- from the 128$^3$ downsampled
  distribution. Vertical lines denote $k_{\rm Ny}/2$ for the three
  cases: 128$^3$ (black), 256$^3$ (blue), 512$^3$ (red). Lower plot:
  ratios of the downsampled power spectra with respect to the 1024$^3$
  particle power spectrum. The dotted line represents the 1$\%$
  deviation limit. }

\vspace{0.5cm}

\centerline{
 \includegraphics[width=3in,clip=]{mass_res_2.eps}}
\caption{\label{fig2}Mass resolution test. The 1024$^3$ particle ICs
  have been downsampled to 512$^3$ and 256$^3$ particles and run to
  $z=0$. The plot shows the ratio of the obtained power spectra at
  $z=1$ (upper panel) and $z=0$ (lower panel) with respect to the full
  1024$^3$ power spectrum at $z=0$ (blue, red) and the power spectra
  which are obtained by downsampling the particles at $z=0$
  (turquoise, orange; see also Figure~\ref{fig1}). The vertical lines
  mark $k=k_{\rm Ny}/2$ for the two cases.}
\end{figure}

We investigate the influence of the particle loading on the accuracy
of the power spectrum by first asking the following question: How many
particles are required to sufficiently sample the density field when
calculating the power spectrum? To answer this question we start from
one of the {\small GADGET-2\/} simulations run with a
($936\,h^{-1}$Mpc)$^3$ box and with 1024$^3$ particles. We determine
the power spectrum from this run at $z=0$. Next, we downsample the
$1024^3$ particles to $512^3$, $256^3$, and $128^3$ particles by
taking the particles which belong to every second (fourth, eight) grid
point in each dimension. Since the particles are downsampled from a
fully evolved simulation, evolution and sampling issues are separated.

In the upper panel of Figure~\ref{fig1} the resulting power spectra
are shown. The lower panel shows the ratio of the power spectra from
the downsampled distributions with respect to the 1024$^3$ particle
distribution. In addition, we have marked the Nyquist wavenumber
divided by two for each power spectrum. The Nyquist wavenumber is set
by the inter-particle separation on the initial grid:
\begin{equation}
\label{nyquist}
k_{\rm Ny}=\frac{\pi}{\Delta_p}=\frac{\pi N_p}{L},
\end{equation}
with $\Delta_p$ being the inter-particle spacing, $N_p$ the cube-root
of the number of particles, and $L$, the box size ($936\,h^{-1}$Mpc)$^3$.
Values of $k_{\rm Ny}$ for the $1024^3$, $512^3$, $256^3$, and $128^3$
particle cases are 3.4, 1.71, 0.86, and $0.43\,h\,{\rm Mpc}^{-1}$,
respectively. As shown in Figure~\ref{fig1}, all power spectra agree
to better than 1\% for $k<k_{\rm Ny}/2$. The undersampled particle
distributions lead to an overprediction of the power spectrum beyond
this point due to the increase in particle shot noise. As mentioned
earlier, a simple shot noise subtraction assuming Poisson noise as
given in Eqn.~(\ref{shot}) does not compensate for this increase.
Detailed tests show that the shot noise which leads to the
overprediction is scale-dependent and smaller than Poisson shot noise
on the scales of interest. (A naive Poisson shot noise subtraction
would alter the power spectrum at $k=1\,h\,{\rm Mpc}^{-1}$ by 0.2\% at
$z=0$ and by 1\% at $z=1$ for 1024$^3$ particles.) Thus we are led to
conclude that, in the absence of shot noise modeling (a difficult and
potentially uncontrolled procedure), the one-percent accuracy
requirement on the power spectrum can only be satisfied for
wavenumbers, $k<k_{\rm Ny}/2$. This quite restrictive limit likely
comes from the fact that the power spectrum is sensitive to
near-mean-density material which is not well modeled on scales smaller
than the mean inter-particle separation.

\begin{figure}
\centerline{
 \includegraphics[width=3in,clip=]{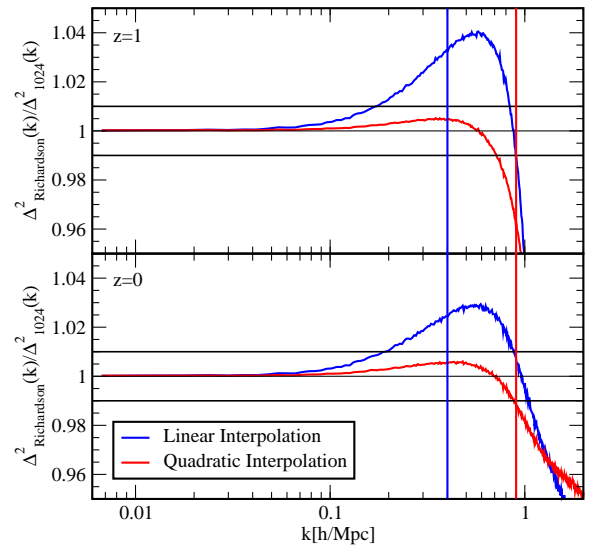}}
\caption{\label{fig3} Richardson extrapolation test. Assuming linear
  [blue curve,~Eqn.~(\ref{estim_lin})] and quadratic convergence [red
  curve,~Eqn.~(\ref{estim_quad})], the 256$^3$ and 512$^3$ particle
  results are used to predict the power spectrum for the 1024$^3$
  particle run. Both plots show the ratio of the prediction with
  respect to the true result. The quadratic extrapolation works well
  in the regime below the half-Nyquist wavenumber, to sub-percent
  accuracy. At $k=1\,h\,{\rm Mpc}^{-1}$ the quality deteriorates due to the
  insufficient resolution of the two underlying runs
  (Cf. Figure~\ref{fig2}). The vertical lines are the same as in
  Figure~\ref{fig2}.}
\end{figure}

The next step is to investigate how the error from an ``undersampled''
initial particle distribution propagates through the numerical
evolution. For this test we first downsample the initial particle
distribution in the same way as before, at $z_{\rm in}=211$, from the
original $1024^3$ particles to $512^3$ particles and $256^3$
particles. We then run the simulations to $z=0$ with the same settings
in {\small GADGET-2\/} as were used for the full run ($2048^3$ PM grid
and a softening length of $50\,$kpc). We do not use the $128^3$
particle set for this test since the corresponding sampling error is
too large. Results are shown in Figure~\ref{fig2} for outputs at $z=1$
and $z=0$. Ratios of the power spectra from the downsampled initial
conditions (ICs) are shown with respect to: (i) the power spectrum
from the full $1024^3$ run, and (ii) the power spectra correspondingly
downsampled at $z=1$ and $z=0$ as shown in Figure~\ref{fig1}.

There are two points to note here. First, restricting attention to
case (i) above, there is a noticeable loss of power below $k_{\rm
  Ny}$, and second, a steep rise beyond this point. The loss of power
is not due to the downsampling in the initial condition -- as can be
easily checked by comparing the power spectrum from the particles
after the IC generation against the desired input power spectrum for
the given realization -- but is due to a discreteness effect: a
reduction in the linear growth factor from its continuum value as
$k\rightarrow k_{\rm Ny}$. As the evolution proceeds, this suppression
is reduced due to the addition of nonlinear power, as can be seen by
comparing the $z=1$ and $z=0$ results in Figure~\ref{fig2}, and also
by noting the smaller suppression for the case with $512^3$ particles
for which the larger $k_{\rm Ny}$ means an enhancement in nonlinearity
(cf. Figure~\ref{fig1}). The steep rise is a manifestation of particle
shot noise as can be seen by looking at the results for case (ii). For
wavenumbers up to $k_{\rm Ny}/2$ there is no difference between the
two ratios [case (i) vs. case (ii)] but beyond that point the results
from case (ii) show a marked reduction ($z=1$) to almost a removal
($z=0$) of the enhancement, consistent with the stated hypothesis. We
would like to re-emphasize that our convergence tests show that a
Poisson shot noise subtraction alters the power spectrum in the wrong
way at the scales of interest. It enhances the suppression of the
power spectrum near the Nyquist wavenumber and overcorrects the power
spectrum at higher wavenumbers.

The problem we now face is that the (IC downsampling) error at $k\sim
k_{\rm Ny}/2$ is large: for the $256^3$ particle run at $z=1$ it is
$\sim 20\%$, and for $512^3$ particles it is still $\sim 7\%$. At
$z=0$, the error is $\sim 10\%$ for the $256^3$ run and $\sim 3\%$ for
the $512^3$ run. Thus, one may wonder if the fiducial $1024^3$
particle run can itself yield results at $k=1\,h\,{\rm Mpc}^{-1}$
accurate to 1\%.

\begin{figure}[t]
\centerline{
 \includegraphics[width=3in,clip=]{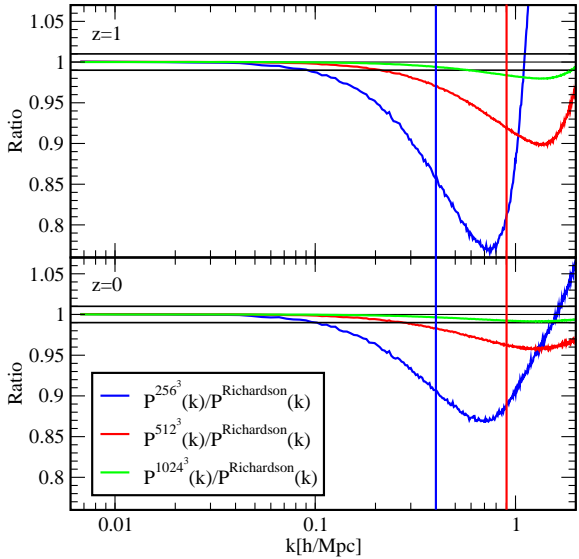}}
\caption{\label{rich_conv} Mass resolution convergence study with
  respect to the extrapolation results at $z=1$ and $z=0$. The content
  of this figure is very similar to that of Figure~\protect\ref{fig3},
  but here we show results with respect to the 2048$^3$ particle
  prediction from the (quadratic) Richardson extrapolation. This
  allows us to investigate the convergence properties of the 1024$^3$
  particle runs (see text for details). The vertical lines are the
  same as in Figure~\ref{fig2}.} 
\end{figure}

A brute force approach would be to run with 2048$^3$ particles and
check convergence with respect to that simulation. To avoid the
computational cost of the brute force approach, we take a different
tack: We extrapolate from the two low-mass resolution runs to try and
predict the results of the high-mass resolution run (see Appendix
\ref{app:rich}). The success of Richardson extrapolation when applied
to power spectra from different force resolution runs has been
demonstrated by \citet{Hei05}. We now carry out a similar procedure,
allowing for both linear or quadratic convergence.

Figure~\ref{fig3} shows the results for the extrapolation tests for
$z=1$ and $z=0$. Following Eqns.~(\ref{estim_lin}) and
(\ref{estim_quad}), we assume linear and quadratic convergence
respectively, and predict the power spectrum for the 1024$^3$ particle
run, displaying the ratio of the prediction with respect to the full
1024$^3$ run. The quadratic extrapolation scheme works much better
than the linear one -- out to $k\simeq 0.8\,h^{-1}$Mpc the prediction
is accurate to better than 1\%. Obviously, the prediction will not
work very well beyond the scale set by the mass resolution of the
256$^3$ simulation. Nevertheless, the test shows that at $k=1\,h\,{\rm
  Mpc}^{-1}$ (which is close to $k_{\rm Ny}/2$ from the $512^3$
particle run and below $k_{\rm Ny}/2$ for the $1024^3$ particle run),
we should obtain a reasonably accurate prediction for a $2048^3$
particle run.

Figure \ref{rich_conv} shows that the $1024^3$ particle run is within
1\% of the prediction for a $2048^3$ run to $k\simeq 1\,h\,{\rm
  Mpc}^{-1}$ at $z=0$ and within 2-3\% at $z=1$ (but here the
extrapolation scheme itself is being stretched to its limit -- the
actual result is likely to be better). This enables us to conclude
that our mass resolution will allow a 1\% accurate calculation at the
scale of interest, without any need to extrapolate.

\subsubsection{Aliasing Effects}

To confirm the results of the tests in this section, we check here for
possible aliasing artifacts which might arise since $N_p\neq N_g$ in
the initial conditions ($N_g$ is the number of grid points per
dimension). We will show briefly in the following that such effects
are negligible.

As explained in Section~\ref{icgen}, the initial conditions in our
simulations are set in the following manner: (i) Implement a
realization of a Gaussian random field initial condition for the
density field in $k$-space, and also for the corresponding scalar
potential and gradients of the potential. (ii) Using an inverse FFT,
determine the gradient field in real space, and use it to move
particles from their initial on-grid positions (where the potential
gradient is exactly known) using the Zel'dovich
approximation. Aliasing cannot enter in the first inverse FFT, but it
can in the second, ``particle move'' step, since the particle grid is
not constrained to be the same as the field grid.

In most simulations, with some exceptions, the typical choice for the
initial condition is to take $\Delta_p=\Delta_g$ or
$\Delta_p=2\Delta_g$ ($\Delta_g$ is the grid spacing) since there is
not much point in adding field power that cannot be represented by the
particle distribution (beyond a spatial frequency set by the particle
Nyquist wavenumber $k_{\rm Ny}$). In addition there is a question that doing
this could be a problem for simulations by leaking artifical ``grid''
power into the initial conditions.

In reality, the situation is relatively benign because of the rapid
fall-off of the initial $P(k)$ at high $k$. This can be seen in
results from earlier papers, e.g., \cite{baugh}, Fig. A.3. Modern
simulations have much higher mass and force resolution, so it is
important to check each time one runs simulations, that there is no
problem with aliased or some other artificial power leaking back to
lower $k$.

The central issue is the existence of the first particle grid peak in
the power spectrum at $k_p=2\pi/\Delta_p$ which influences the
computation of $P(k)$ close to it in a way that is hard to correct or
compensate for, given that we are interested in percent level
accuracy. For a chosen $k$ scale of interest, $k_I <k_{\rm Ny}$, one
has to make sure that $k_p$ is sufficiently greater than $k_I$ at the
redshift of interest [the lower the redshift the easier to satisfy
this condition, since evolution boosts $P(k_I)$ significantly compared
to $P(k_p)$].

\begin{figure}[t]
\centerline{
 \includegraphics[width=2.4in,angle=270]{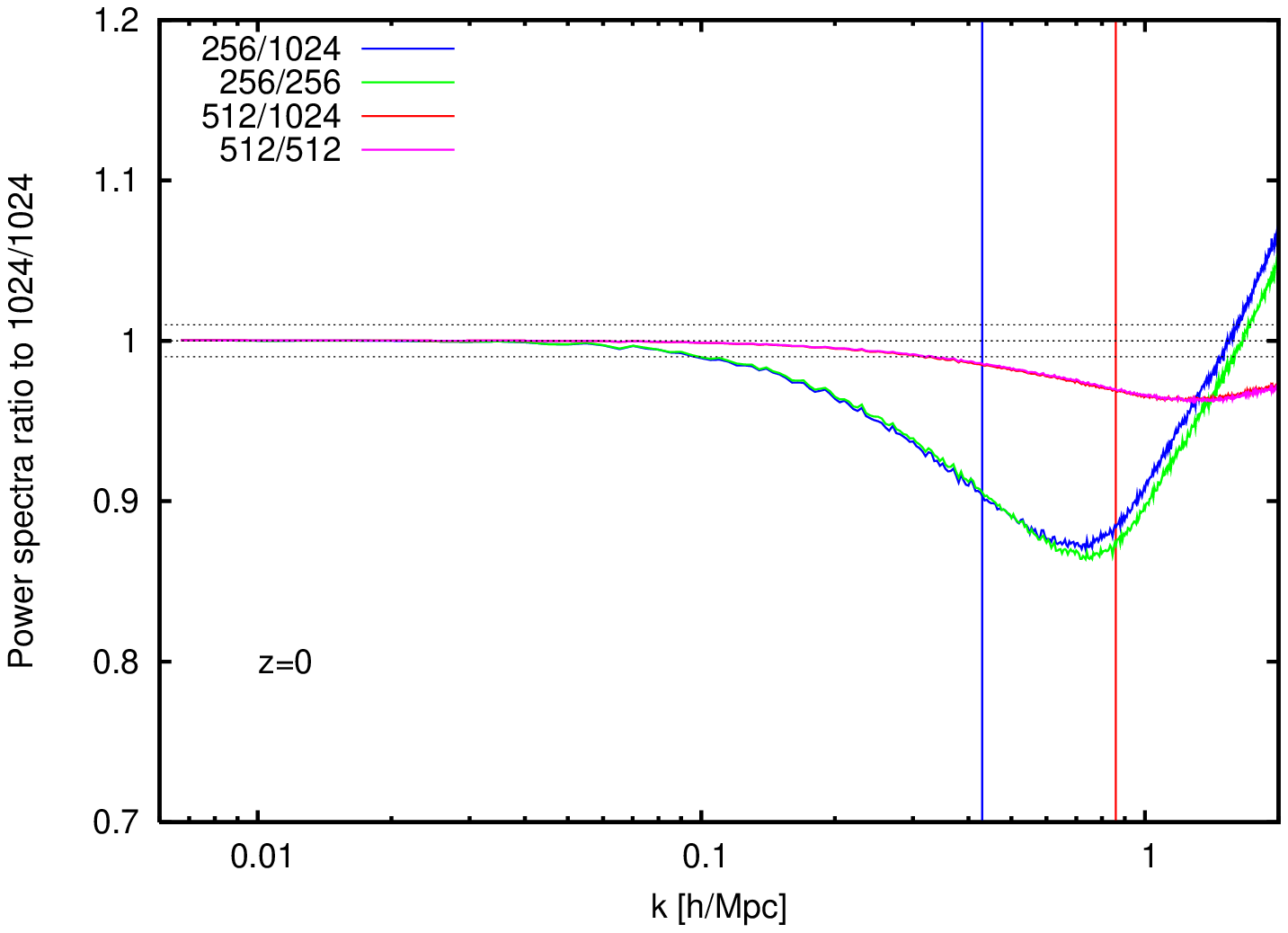}}
\caption{Ratio of power spectra showing insensitivity to the presence
  of $k>k_{\rm Ny}$ power in the initial conditions. Note the
  excellent agreement between the 256/256 and 256/1024 runs and the
  512/512 and 512/1024 runs. The vertical lines denote $k_{\rm Ny}/2$
  for the two cases.}
\label{alias}
\end{figure}

In the specific mass resolution tests carried out above we investigate
the case of a single realization with fixed $\Delta_g$ for different
choices of $\Delta_p$. In order to show that potential aliasing
effects do not alter our results we carry out the following additional
test. We fix $N_g=1024$ and consider two cases with $N_p=512$ and
$N_p=256$ (corresponding to $\Delta_p=2\Delta_g$ and
$\Delta_p=4\Delta_g$). In addition to these runs we also run three
simulations all with $\Delta_p=\Delta_g$ with $N_p=1024$, $N_p=512$,
and $N_p=256$, explicitly setting all the high-$k$ modes to zero for
the latter two cases, for the same $k$ space realization as in the
first. Thus we have essentially the same phases but no power beyond
$k_{\rm Ny}$ in all three cases. The results for $P(k)$ are shown at
$z=0$ as a ratio against the $N_g=N_p=1024$ case in
Figure~\ref{alias}. Note that the same suppression of power around
$k_{\rm Ny}$ as noted in the previous section is seen here,
independent of whether high $k$ power is present in the initial
conditions or not. Thus any effect due to aliasing is negligible.

\subsection{Force Resolution}

\begin{figure}[t]
\centerline{
 \includegraphics[width=3in,clip=]{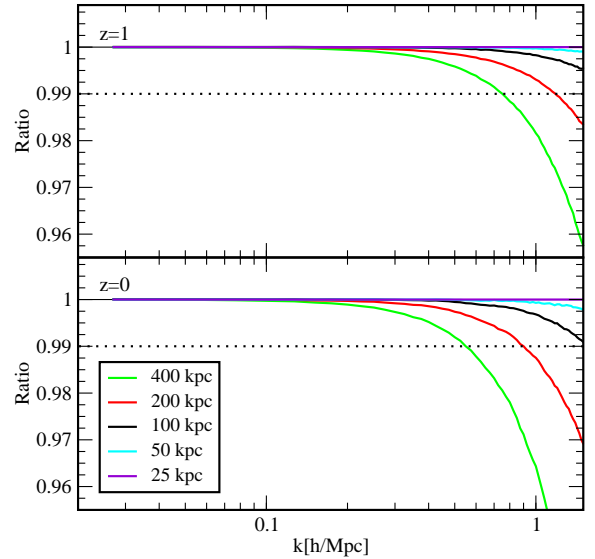}}
\caption{Force resolution convergence study at $z=1$ and $z=0$ with
{\small GADGET-2}. The 512$^3$ PM grid is the same in all five runs,
and the force resolution is varied between 25~kpc and 400~kpc. At
$k\sim 1\,h\,{\rm Mpc}^{-1}$, a force resolution of 100~kpc already
leads to results converged well below 1\% at both redshifts with
respect to the 25~kpc resolution run.}
\label{force_conv}
\end{figure}

As discussed in Section~\ref{nbody} we employ two $N$-body methods in
this paper: PM simulations with grid sizes of 1024$^3$ and 2048$^3$
and tree-PM simulations. The force resolution of the PM runs is
insufficient to resolve the power spectrum out to $k\sim 1\,h\,{\rm
  Mpc}^{-1}$ (see, e.g., Figure~\ref{match} for the shortfall of power
in the PM runs). We therefore discuss only the convergence properties
of the tree-PM algorithm out to $k~\sim 1\,h\,{\rm Mpc}^{-1}$. Since
the {\small GADGET-2\/} runs with $1024^3$ particles are
computationally expensive, and the force softening primarily affects
small scales, we chose to downscale the simulation box and number of
particles for this test to $256^3$ particles in a $234\,h^{-1}$Mpc box
(a reduction by a factor of 64 from the main runs). Following the
practice in the larger runs, the PM force grid is set to twice the
number of particles in one dimension, resulting in a $512^3$ PM mesh.
All the other code settings are the same as for the large runs and we
vary only the force softening to test for the effects of finite force
resolution. The effective force resolution lengths range from
$400\,$kpc to $25\,$kpc ($50\,$kpc is used in the large runs). The
results for $z=0$ and $z=1$ are shown in Figure~\ref{force_conv}. At
$k\sim 1\,h\,{\rm Mpc}^{-1}$, the difference between $50\,$kpc and
$25\,$kpc is well below 0.1\% for both redshifts, and therefore
comfortably within our requirements. In fact, meeting the force
resolution requirements at $k\sim 1\,h\,{\rm Mpc}^{-1}$ with the
tree-PM algorithm is computationally much less demanding than meeting
the mass resolution requirements. It may be that for power spectrum
simulations a hybrid or adaptive PM code is the most computationally
efficient route, though other uses of the simulations may be more
sensitive to resolution.

The size of the PM mesh is a separate issue, and significant in its
own right. If high accuracy is desired the mesh should not be chosen
to be too small, as this increases the PM error and pushes the
handover between the tree and the mesh to larger scales. In tests
carried out to determine the size of the PM grid, we observed an
unphysical suppression of the early-time power spectrum at
quasi-linear scales for the smaller meshes.

\subsection{Time Stepping}

Most $N$-body codes use low-order -- typically, second order --
symplectic time-stepping schemes. (Full symplecticity is not achieved
when adaptive time-stepping is employed.) The choice of the time
variable itself can vary, although typically it is some function of
the scale factor $a$, e.g., $a$ itself or the natural logarithm of
$a$. PM codes most often use constant time stepping in $a$ or $\ln a$.
Higher-resolution codes use adaptive, as well as individual particle
time-stepping. Hybrid codes that mix grid and particle forces, such as
tree-PM, have different criteria for time-stepping the long-range
forces as compared to the short-range forces, where individual
particle time-steps are often used. Because of these complexities, it
is important to check that the time-stepping errors are sub-dominant
at the length-scales of interest for computing the mass power
spectrum.

\begin{figure}[t]
\centerline{
 \includegraphics[width=3in,clip=]{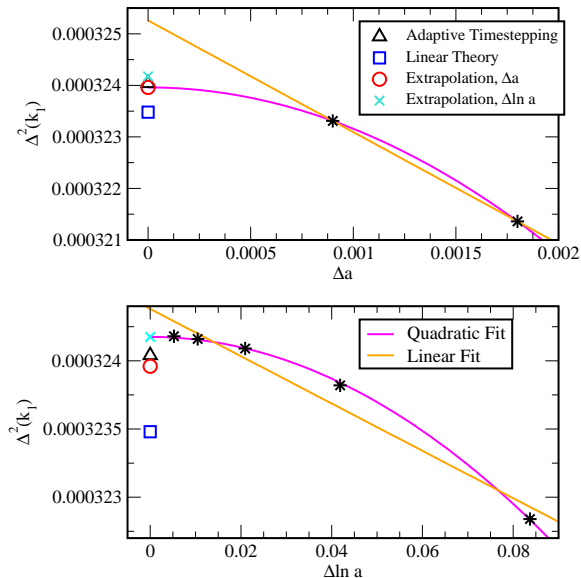}}
\caption{Time stepper convergence for $\Delta^2(k)$ using linear (upper
  plot) and logarithmic (lower plot) time-stepping in $a$ at $z=0$, as
  a function of number of time steps. The $k$ value chosen is for the
  largest mode in the box, $k=6.7\times 10^{-3}\,h\,{\rm Mpc}^{-1}$ 
  (black stars). The black triangle shows the result from a {\small
    GADGET-2} run with adaptive time stepping in $\ln a$, the blue box
  is the power spectrum from the initial condition scaled by the
  linear growth factor to $z=0$, the red circle the $\Delta^2(k)$ value for
  time-stepping linear in $a$ extrapolated to zero assuming quadratic
  convergence, and the turquoise cross the same quantity for $\ln
  a$. All (extrapolated) values from the simulations agree with linear
  theory to 0.2\% or better, the simulations themselves agreeing to
  better than 0.04\% taking the {\small GADGET-2} run as the
  reference. The pink line shows a quadratic fit to the data points.
  \label{timestep}}
\end{figure}

The {\small GADGET-2\/} runs in this paper use $\ln a$ as the time
variable. The PM calculations within {\small GADGET-2\/} use a global
time step; we found 256 time steps sufficient for this part. The tree
algorithm for the short-range forces uses an adaptive time stepping
scheme and our runs use a total of about 3000 time steps. The
criterion for the adaptive time stepping is coupled to the softening
length $\epsilon$ via : $\Delta t = \sqrt{2\eta\epsilon/|a|}$ where
$\eta$ allows adjustments in the time stepping; we use $\eta = 1\%$
(note that here $a$ is the acceleration). Detailed tests of the
convergence of the time stepping employed by {\small GADGET-2\/} can
be found in Section 4 of \cite{Spr05}.

We perform an additional test to verify the expected quadratic
convergence, considering the largest mode in the box (in this case
$k=6.7\times 10^{-3}\,h\,{\rm Mpc}^{-1}$). We compare the numerical
results for $P(k)$ with that expected from linear theory, which should
be reasonably accurate at these very large scales. By using the
largest mode, one is insulated from errors due to the particle loading
and small-scale force resolution.

We investigate both time variable choices, $\ln a$ and $a$. The
results are shown in Figure \ref{timestep}. All the test runs are in
pure PM mode on a $1024^3$ grid, with the tree switched off in {\small
  GADGET-2\/} (there is no need for high force resolution in this
test) and using global time-stepping. For steps linear in $a$ we show
results for roughly 600 and 1,200 time steps, for the time stepper in
$\ln a$ we show results for $\Delta \ln a \approx$ 0.005, 0.01, 0.02,
0.04, and 0.08. In addition, we fit two curves through the results
assuming linear and quadratic convergence. As expected from a second
order integrator, the quadratic fit is in very good agreement with the
data points. Quadratic extrapolation of the results for the two time
stepping schemes from finite $k$ to zero is in very good agreement
with linear theory, to better than 0.2\% -- about the deviation
expected given the dimensionless power at the fundamental mode of the
box. If we take the adaptive time step run as the reference (rather
than linear theory), the agreement is better than 0.04\%. Adaptive
time-stepping is expected to yield results very close to $\ln a$
stepping on large scales, since for the long-range force even the
adaptive time-stepper run is constant in $\ln a$ with $\Delta\ln a
=0.02$. The excellent agreement with time-stepping in $a$ confirms the
robustness of the different schemes. Since our interest is in
generating the power spectrum at percent accuracy at minimal computing
cost, we conclude that the $\ln a$ time-stepping scheme with
approximately 250 time steps is a good compromise for the PM runs to
obtain an accurate power spectrum at quasi-linear scales (two orders
of magnitude removed from scale set by the force resolution).

\section{Matching low and high resolution power spectra and 
comparison with {\sc HaloFit}}
\label{sec:halofit}

\begin{figure}[b]
\centerline{
 \includegraphics[width=3in]{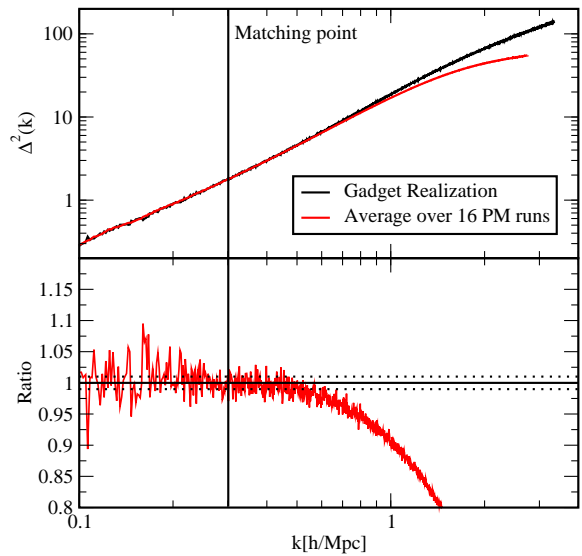}}
\caption{\label{match}Matching of an ensemble of low resolution runs
  with one realization of a high-resolution {\small GADGET-2} run. The
  upper panel shows the average from 16 realizations from the low
  resolution PM runs (red) and the power spectrum from the {\small
    GADGET-2} run (black). The lower panel shows the ratio of the low
  resolution ensemble with respect to the {\small GADGET-2} run. Out
  to $k\sim0.5\,h\,{\rm Mpc}^{-1}$ the difference is less than one percent
  (disregarding the noise from the single realization). We match the
  two power spectra at $k\sim0.3\,h\,{\rm Mpc}^{-1}$, at which point the
  noise in the single realization is small enough, yet the resolution
  of the PM runs is sufficient to accurately resolve the power
  spectrum.}
\end{figure}

Last, we compare our results with the standard fitting formula, {\sc
  HaloFit\/} \citep{Smi03}, currently used for analysis of e.g. weak
lensing data \citep{Jar06,Mas07,Ben07,Fu08} or for forecasts on the
improvement of cosmological constraints from future surveys
\citep{Tan08}. {\sc HaloFit\/} provides the nonlinear power spectrum
over a range of cosmologies in a semi-analytic form. It is based on a
combination of the halo model approach (for a review of the halo
model, see e.g., \citealt{HaloModel}) and an analytic description of
the evolution of clustering proposed by \citet{Ham91}. In addition,
the fit is tuned to simulations by introducing two new parameters: an
effective spectral index on nonlinear scales, $n_{\rm eff}$, and a
spectral curvature $C$. The combination of analytic arguments and
tuning to results from $N$-body simulations has led to the most
accurate fit for the nonlinear power spectrum to date (as we will show
below, the fit is accurate to $\sim 5-10$\%). As mentioned above, we
use here the CAMB implementation of {\sc HaloFit}.

\begin{figure}[t]
\centerline{
 \includegraphics[width=3in]{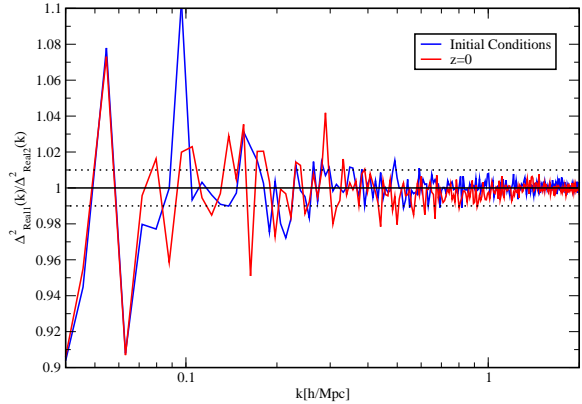}}
\caption{\label{real} Ratio of power spectra from two independent
  realizations at initial and final redshifts. Both simulations are
  carried out with {\small GADGET-2} at the standard setting. Results
  as shown have been smoothed by  averaging over every five
  $k$-values. Beyond our matching point for low and high resolution
  simulations,  $k=0.3$~$h$Mpc$^{-1}$, the results agree at the
  percent level, confirming that one realization of an $\sim h^{-1}$Gpc
  high-resolution run is sufficient.}
\end{figure}

\begin{figure}[t]
\centerline{
 \includegraphics[width=3in]{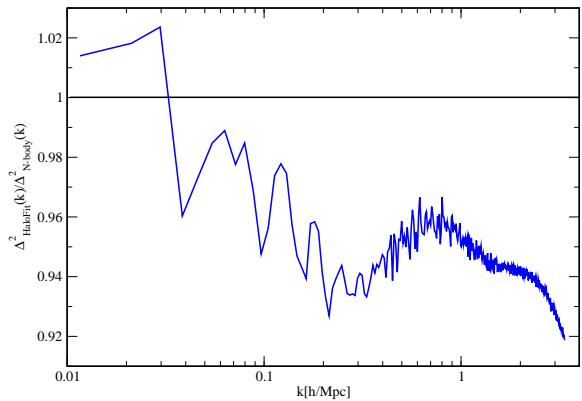}}
\caption{\label{halofit}Comparison of the simulation power spectrum to 
  {\sc HaloFit}. Shown is the ratio of {\sc HaloFit\/} with respect to
  the simulation result. The simulation result has been obtained by
  combining the PM runs and the {\small GADGET-2} run at
  $k=0.3\,h\,{\rm Mpc}^{-1}$ and it has been smoothed by averaging
  over every five $k$-values to reduce the noise for the comparison.
  The {\sc HaloFit} result is approximately 5\% lower than the result
  from simulations.}
\end{figure}

In order to compare simulation results to a smooth fit, we first
combine 16 realizations from the PM runs in the low $k$ region with
one high-resolution run, as shown in Figure~\ref{match}. At around
$k=0.6\,h\,{\rm Mpc}^{-1}$ the lower resolution of the PM runs begins
to become apparent and the result falls below that from {\small
  GADGET-2}. Conservatively, we match the two power spectra at
$k=0.3\,h\,{\rm Mpc}^{-1}$. At this point, the variance from the
single realization of the {\small GADGET-2\/} run is small enough that
the matching leads to a smooth power spectrum. (A more sophisticated
matching procedure is described in \citealt{Law09}, Paper III.) One
concern might be that a single realization is insufficient to capture
the behavior on small scales accurately: Because of mode coupling it
is not obvious that fluctuations on large scales do not also cause
substantial effects on small scales. In Figure~\ref{real} we show
that, due to the large box size, this is not a concern at least at the
percent level of accuracy. The figure shows the ratio of two different
realizations at the initial and final redshift. Both simulations are
run with {\small GADGET-2\/} at our standard settings. The variations
at high $k$ (beyond the matching point $k=0.3\,h\,{\rm Mpc}^{-1}$ )
are at the percent level and appear to be free of systematic trends.

The ratio of the matched power spectrum to the prediction from {\sc
  HaloFit} is shown in Figure~\ref{halofit}. In this case, the {\sc
  HaloFit\/} prediction falls roughly 5\% below the simulation. The
procedure for combining the simulation results can be seen to work
very well, as there is no discontinuity at $k=0.3~h$Mpc$^{-1}$ from
the matching. Our result is in good agreement with, e.g.,
\citet{Smi08} as well as \citet{Ma07}, who find a 5\% supression for
{\sc HaloFit\/} at $k\sim 0.1~h$Mpc$^{-1}$. At larger $k$, however,
the results in \citet{Ma07} may not be very accurate, due to
limitations in force resolution in that work.

\section{Conclusion and Outlook}
\label{conclusion}
 
The advent of precision cosmological observations poses a major
challenge to computational cosmology. With observational results
accurate to the percent level a significant uncertainty in extracting
cosmological information from the data is due to inaccuracies in
theoretical templates. At the required level of accuracy large scale
simulations are unavoidable, since the nonlinear nature of the problem
makes it impossible to derive analytic or semi-analytic expressions
for statistics such as the matter power spectrum, at an accuracy
better than $\sim 10\%$. While simulations in principle should yield
results at sub-percent accuracy, in practice this is a non-trivial
task due to uncertainties in the numerical implementation and modeling
of relevant physical processes.

Motivated by this realization, we decided to carry out an end-to-end
calculation of one of the simplest non-trivial problems we could
imagine: a percent level computation of the nonlinear mass power
spectrum to $k\sim 1\,h\,{\rm Mpc}^{-1}$ over the range $0<z<1$. This
was a problem which appeared useful and timely as well as tractable
(if not straightforward) while still providing a meaningful learning
environment -- by actually going through all of the steps we would map
out the necessary infrastructure which would be required, find the
most difficult pieces of the problem and present a proof-of-principle
demonstration that meaningful, precision theoretical predictions could
be used in support of future cosmological measurements.

We have broken the problem into three steps, to be presented in three
publications. In this, first, paper we showed that it is possible to
obtain a calibration of the nonlinear matter power spectrum at
sub-percent/percent accuracy out to $k\sim 1\,h\,{\rm Mpc}^{-1}$
between $z=1$ and $z=0$. This wavelength regime is important for
ongoing and near-future weak-lensing surveys. The restriction to these
(large) length scales has two major advantages: baryonic effects are
subdominant on these scales
\citep[e.g.][]{Whi04,ZhaKno04,Jin06,Rud08,GuiTeyCol09} and the
numerical requirements in this regime remain rather modest. Each
simulation can be carried out in a matter of days on parallel
computers with several hundred processors and the data volume is
manageable with arrays of inexpensive disk. Pushing beyond $k\sim
1\,h\,{\rm Mpc}^{-1}$ will require advances in our understanding of
the implementation of baryonic physics, or self-calibration
techniques, as well as advances in algorithms and computational power.

We derived a set of numerical requirements to obtain an accurate power
spectrum by performing a large suite of convergence and comparison
tests. The goal was a set of code settings which balance the need for
precision and the limitation of computational resources. As shown
here, the simulation volume and, especially, the particle loading are
two major concerns in obtaining an accurate matter power spectrum. The
simulation volume has to be in the $\sim$Gpc$^3$ range, leading to a
minimum requirement of $\sim 1$ billion particles. Further increase in
volume would be helpful, but would require a concomitant increase in
the number of particles, greatly adding to the computational burden.
The 1~Gpc$^3$/1~billion particle simulation is a good compromise
between sufficient accuracy and computational cost.

Besides a large simulation volume and good particle sampling,
initialization of the simulation also plays an important role. To
guarantee converged results, the simulation must be started at a high
enough redshift. We found that a starting redshift of $z_{\rm
  in}\simeq 200$ is sufficient to get accurate results between $z=1$
and $z=0$.

The results for the power spectrum are rather stable to changes in the
number of time steps. This is clearly related to the fact that our
resolution demands are relatively modest. For the PM runs, a few
hundred time steps are sufficient, while for the tree-PM runs the
overall number of time steps is a factor of ten larger. We emphasize
that the simulation settings discussed here will lead to the required
accuracy only up to $k\sim1\,h\,{\rm Mpc}^{-1}$. While these settings
can be used as a guideline for other simulation aims, they do not
replace convergence tests that must be performed for each new problem,
if one desires high precision results.

While weak lensing was a primary motivation for this study, our
efforts are of wider interest as an exercise in precision
``theoretical'' cosmology.  We demonstrated that it is possible to
achieve 1\% accuracy in the mass power spectrum in gravity only
simulations on relatively large scales for a limited range of
cosmological models.  Had this not been the case the field would have
needed to rethink its demands on theory.  The non-trivial
computational and human cost of even this ``first step'' argues for
increased efforts in these directions in order to satisfy the
increasingly stringent demands of future observations.

Having established the ability to generate power spectra with
sufficient accuracy from $N$-body simulations, the next major question
that arises is how to use these costly simulations for parameter
estimation, e.g., via Markov Chain Monte Carlo. To address this
problem, we have recently introduced the cosmic calibration framework
\citep{Hei06,Hab07,Sch08} which is based on an interpolation scheme
for the power spectrum (or any other statistic of interest) derived
from a relatively small number of training runs.

The next step in generating precise predictions for the matter power
spectrum is to determine the minimum number of cosmological models
needed to build an accurate emulator and then to construct the
emulator from a set of high-precision simulations.  In the second
paper of this series we establish that 30-40 cosmological models are
sufficient to explore the parameter space for $w$CDM cosmologies
(constant $w$) given the current constraints on parameter values.  The
third and final paper will present results from the simulation suite
designed and discussed in the second paper, and will include a power
spectrum emulator that will be publicly released.

\acknowledgements

A special acknowledgment is due to supercomputing time awarded to us
under the LANL Institutional Computing program and from the LRZ Munich
under the German AstroGrid Initiative. Part of this research was
supported by the DOE under contract W-7405-ENG-36. SH, KH, DH, EL, and
CW acknowledge support from the LDRD program at Los Alamos National
Laboratory. MJW was supported in part by NASA and the DOE. We would
like to thank Nikhil Padmanabhan for useful discussions, Stefan
Gottl\"ober and Anatoly Klypin for help with the ART comparison, and
Volker Springel for making {\small GADGET-2\/} publicly available.

\appendix

\section{Convergence Tests for Initial Conditions}
\label{app:2lpt}

\begin{figure}[b]
\begin{center}
\resizebox{3in}{!}{\includegraphics{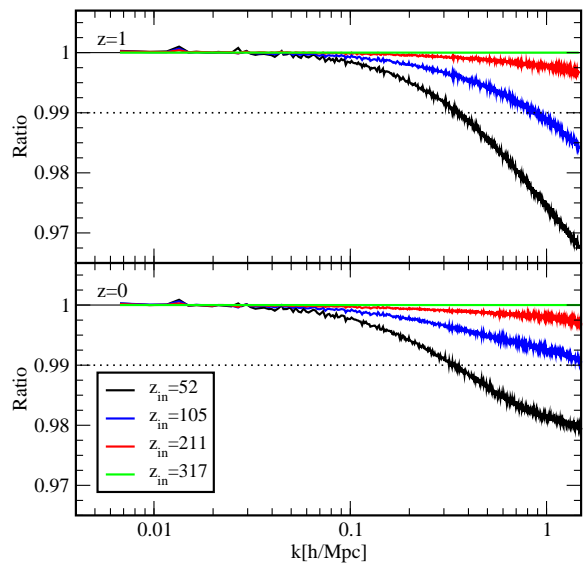}}
\end{center}
\caption{Comparison of ratios of the dimensionless power 
spectra at $z=1$ (upper panel) and $z=0$ (lower panel) when evolved
using a PM code from initial conditions generated using the Zel'dovich
approximation at the starting redshifts indicated. The rms
displacement for the starts is 0.335, 0.168, 0.084, and 0.055 times the
mean inter-particle spacing (for $z_{\rm in}=52$, 105, 211 and
317). The dotted lines mark the 1\% limit. If the code is started at
$z_{\rm in}=52$, we see a suppression of the power spectrum by $\sim
3\%$ at $z=1$ and $\sim 2\%$ at $z=0$. }
\label{fig:ic_z0}
\end{figure}

The initial conditions for $N$-body simulations are usually generated
by displacing particles from a regular grid using the Zel'dovich
approximation. This amounts to a first order expansion in Lagrangian
perturbation theory. In order to verify that our criteria for the
initial redshift, explained in Section \ref{ics}, are sufficient to
guarantee one percent accuracy between $z=1$ and $z=0$ we carry out a
convergence study.

The first step is indicated in Figure~\ref{fig:ic_z0}, which shows
that the power spectrum between $z=1$ and 0 converges as we increase
$z_{\rm in}$ and is well converged by $z=0$ given $z_{\rm in}$
satisfying our criteria. Our results are in very good agreement with
similar tests carried out by, e.g., \citet{Ma07}. We carried out
numerous other tests with very similar results including tests for
different cosmologies. By starting when $D(z_{\rm in})/D(z=1)=0.01$
our results are converged to better than $1\%$ for all $0\le z\le 1$.

\begin{figure}[t]
\begin{center}
\resizebox{2.5in}{!}{\includegraphics{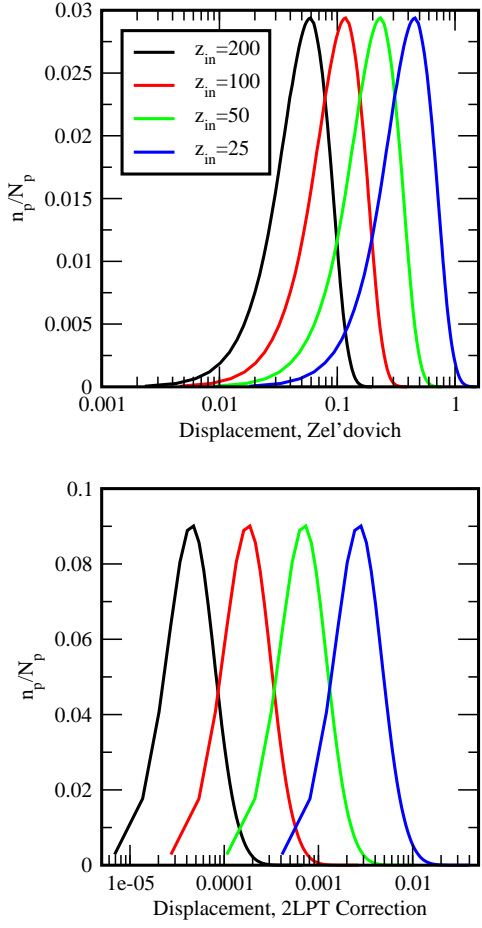}}
\end{center}
\caption{Upper panel: distribution of the initial displacements of all
  particles at different starting redshifts ($z_{\rm in}=200,~100,~50,~25)$.
  The displacement is measured with respect
  to the mean inter-particle spacing. For $z_{\rm in}=200$, the rms
  displacement is approximately 0.05, while for $z_{\rm in}=25$ it
  increases by a factor of ten. Lower plot: 2LPT correction. The
  distributions show the additional contribution in the initial move
  to the Zel'dovich approximation. For $z_{\rm in}=200$ this
  additional move is on average $4\cdot 10^{-5}$ and for $z_{\rm
    in}=25$ it is 0.004 of the mean inter-particle spacing. In both
  cases this is a small fraction with respect to the Zel'dovich
  move. In both plots, the $y$-axis is scaled with respect to all
  particles.}
\label{fig:disp}
\end{figure}

\begin{figure}[t]
\begin{center}
\resizebox{2.5in}{!}{\includegraphics{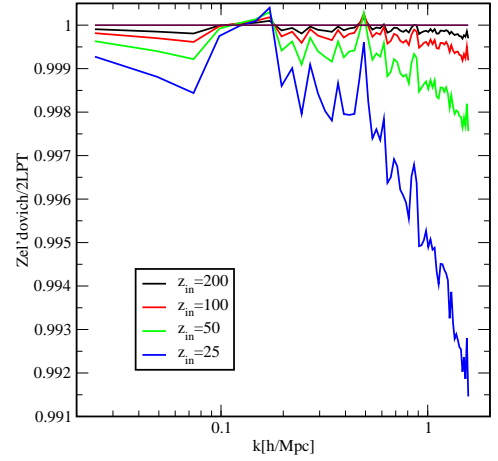}}
\end{center}
\caption{Power spectra ratios for four different initial
  redshifts. The initial power spectrum obtained from Zel'dovich
  initial conditions is divided by the power spectrum from the 2LPT
  initial conditions. Overall, the Zel'dovich initial conditions have
  slightly less power on the smallest scales. Results are shown out to
  $k_{\rm Ny}/2=\pi N_p/2L=1.57\,h\,{\rm Mpc}^{-1}$.  Remarkably, even
  if the initial conditions are generated as late as $z_{\rm in}=25$,
  the difference in the power spectra is below 1\% at the smallest
  scales. For $z_{\rm in}=200$, the difference on all scales is far
  below one percent. \cite{nish08} found a similar result: sub-percent
  agreement between power spectra from the Zel'dovich approximation
  and 2LPT initial conditions at $z=127$.}
\label{fig:pow}
\end{figure}

The second step is to show that the results as $z_{\rm in}\to\infty$ are
converging to the desired answer.  One way to check this is to compare the
ZA scheme to a higher order Lagrangian approximation, e.g.~$2^{\rm
  nd}$ order Lagrangian perturbation theory: 2LPT. (The use of a
higher order Lagrangian approximation scheme to set up initial
conditions has been suggested recently, e.g., \citealt{Cro06}.) For
small initial perturbations 2LPT should be more accurate than ZA, and
generates transients which decay much faster with the expansion of the
Universe ($a^{-2}$ rather than $a^{-1}$). In the 2LPT formalism, the
particle displacement is obtained in second order Lagrangian
perturbation theory, an additional contribution being added to that
from the Zel'dovich approximation as given in Eqn.~(\ref{zeldo}):
\begin{eqnarray}
\label{disp}
\bf x(\bf q)&=&{\bf q}-D_1{\bf \nabla}_q\phi^{(1)} +
D_2{\bf\nabla}_q\phi^{(2)},\\ 
\bf v&=&\frac{d{\bf x}}{dt}=-D_1f_1H{\bf
\nabla}_q\phi^{(1)}+D_2f_2H{\bf \nabla}_q\phi^{(2)},
\end{eqnarray}
where $\phi^{(2)}$ is obtained from solving 
\begin{equation}
{\bf \nabla}^2_q\phi^{(2)}({\bf
q})=\sum_{i>j}\left\{\phi_{,ij}^{(1)}({\bf q})\phi^{(1)}_{,ij}({\bf
q})-[\phi^{(1)}_{,ij}({\bf q})]^2\right\} 
\end{equation}
and $D_2$ is the second order growth function. In the following, we
investigate the contributions from the second terms in the positions
and velocities of the particles at different redshifts.

\begin{figure}[t]
\begin{center}
\resizebox{3.in}{!}{\includegraphics{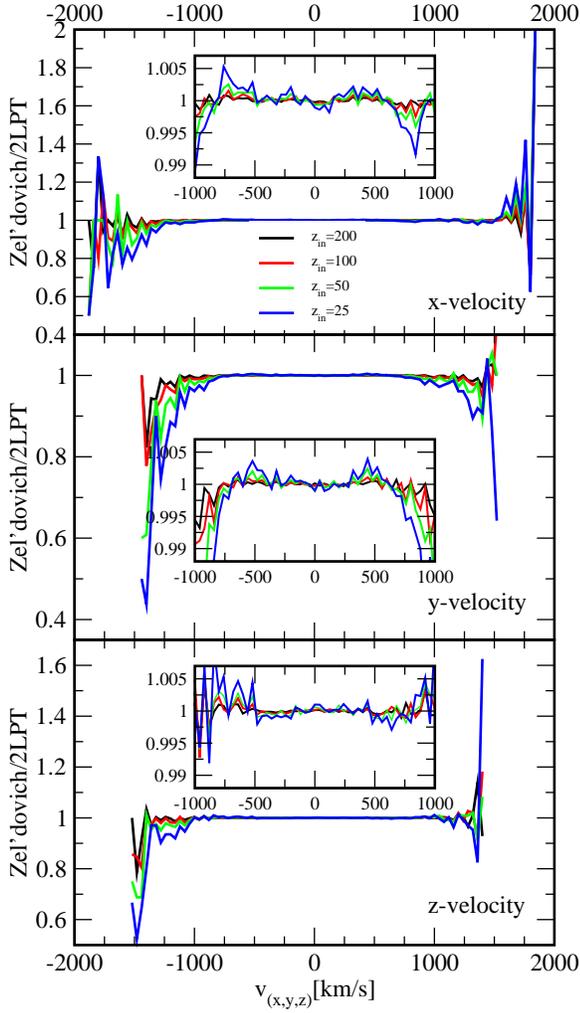}}
\end{center}
\caption{Ratio of histograms of the three velocity components from the
  Zel'dovich approximation and the 2LPT approach. The insets show the regimes
  between -1000~km/s and 1000~km/s where the large majority of the
  particles reside. Here the difference is sub percent. The different
  colors represent different starting redshifts, the difference
  becoming smaller for higher redshift starts.}
\label{fig:velc}
\end{figure}

\cite{Cro06} have made a serial 2LPT code publicly available. Their
code uses approximations for the growth functions in first and second
order. (In contrast, the ZA initialization routine used for this paper
solves the differential equation for the linear growth function
directly, without making approximations.) For a $\Lambda$CDM cosmology
these approximations are given by:
\begin{eqnarray}
D_1&\approx& \frac{5}{2} a
\Omega_m\left[\Omega_m^{4/7}-\Omega_\Lambda+
  \left(1+\frac{\Omega_m}{2}\right)\left(1+ 
\frac{\Omega_\Lambda}{70}\right)\right]^{-1},\\ 
D_2(\tau)&\approx& \frac{3}{7}D_1^2(\tau) \Omega_m^{-1/143}\approx
-\frac{3}{7}D^2_1(\tau), 
\end{eqnarray}
with $\tau$ being conformal time. The approximation for $D_1$ can 
be found in \citet{Car92}. For $f_1$ and $f_2$ the following
approximations are made:
\begin{equation}
f_1\approx \Omega_m^{5/9},~~~f_2\approx 2\Omega_m^{6/11}.
\end{equation}
A detailed discussion of the exact differential equations for the
growth function up to third order and the reliability of these
approximations is given in \citet{Bou95}. In order to limit
computational expense, we restrict our tests using this code to
256$^3$ particles in a $256\,h^{-1}$Mpc volume. This choice is
sufficient to study the general question, as the inter-particle
spacing is the same as in the main runs.  In keeping with our general
philosophy of redundancy and cross-checking we also independently
implemented a 2LPT initial conditions generator (with numerical
computation of the growth functions, rather than approximations) which
gave essentially the same results as that of \citet{Cro06}.

\begin{figure}[t]
\begin{center}
\resizebox{2.5in}{!}{\includegraphics{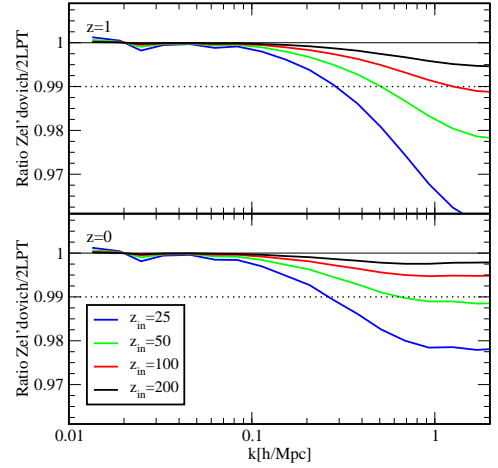}}
\end{center}
\caption{Comparison of the power spectra from simulations started from
Zel'dovich initial conditions and 2LPT initial conditions at $z=1$
(upper panel) and $z=0$ (lower panel). Shown are the ratios of power
spectra from starts at redshift $1+z_{\rm in}=200, 100, 50, 25$. The
start at $1+z_{\rm in}=200$ leads to an agreement of the power spectra
better than 0.5\% at $k\sim 1~h$Mpc$^{-1}$ and better than 0.2\% at
$z=0$.  At larger scales, $k<0.1\,h\,{\rm Mpc}^{-1}$, the agreement is
basically perfect.  Therefore, the Zel'dovich initialization scheme
started at $1+z_{\rm in}=200$ fulfills our accuracy requirements
comfortably.}
\label{fig:powevol}
\end{figure}

We generate four sets of initial conditions at $z_{\rm in}=200$, 100,
50 and 25.  All of the initial conditions have the same phases and can
therefore be compared directly.  First, we measure the displacement
from the Zel'dovich approximation; results are shown in the upper
panel of Figure \ref{fig:disp}.  For this one realization, the rms
displacement at $z_{\rm in}=200$, which is the starting redshift for
our main simulations, is around 5\% of the mean interparicle spacing.
By delaying the start until $z_{\rm in}=25$, the rms displacement
grows by a factor of ten.  The 2LPT correction, given by the second
term in Eqn.~(\ref{disp}), is negligible at $z_{\rm in}=200$, being
smaller than $10^{-4}$ on average.  In fact at this point numerical
accuracy might be questioned, since the approximations for the growth
functions might not be accurate at this level.  Figure~\ref{fig:pow}
shows the ratio of the initial power spectra from the Zel'dovich and
the 2LPT approximations.  As for the displacements, convergence with
increased redshift is very apparent.  At a starting redshift of
$z_{\rm in}=200$, both power spectra agree to better than 0.02\%.
Even starting at very late times ($z_{\rm in}=25$) only leads to a 1\%
difference between the initial power spectra.

Next we measure the differences in the initial velocities from the two
approximations.  The results are shown in Figure \ref{fig:velc}.  We
display the three velocity components $v_x$, $v_y$, and $v_z$
separately.  The main difference occurs in the tails of the velocity
distributions.  Independent of redshift a negligible number of
particles (fewer than 0.5\%) live in these tails with absolute initial
velocities larger than $1000\,$km/s.  Ignoring these tails (see the
insets in Figure \ref{fig:velc}), the difference in the velocities
between 2LPT and ZA starting at different redshifts is below 1\%.  At
$z_{\rm in}=200$ the difference is less than 0.1\%.  At this
precision, the inaccuracy from the approximations for the growth
function at first and second order is probably larger than the error
from the Zel'dovich approximation.

The velocity differences are highly correlated with density however
(see also Figure \ref{fig:vis}), and to understand this effect we
evolve initial conditions created from the ZA and 2LPT forward to
$z=0$.  We use our parallel 2LPT code, which does not rely on an
approximation for the growth function, to generate initial conditions
with $512^3$ particles in a $468\,h^{-1}$Mpc box -- downscaling our
main runs by a factor of eight.  The ICs are generated at four
different redshifts, $1+z_{\rm in}=200$, 100, 50 and 25 and evolved to
$z=0$ using a tree-PM code.  We measure the power spectrum of the
evolved particles at $z=1$ and $z=0$.  The results are shown in
Figure~\ref{fig:powevol}, where we see a shortfall in power at high
$k$ in the ZA starts as compared to the 2LPT starts but convergence as
$z_{\rm in}$ is increased.  At $k\sim 1\,h\,{\rm Mpc}^{-1}$ the
evolved power spectra from both sets of initial conditions at
$1+z_{\rm in}=200$ show excellent agreement, better than 0.5\% at
$z=1$ and 0.25\% at $z=0$.  We therefore conclude that our starting
redshift, $1+z_{\rm in}=200$, is high enough to avoid any problems
arising from possible inadequacies of the Zel'dovich approximation.

\begin{figure}[t]
\begin{center}
\resizebox{2.6in}{!}{\includegraphics[angle=270]{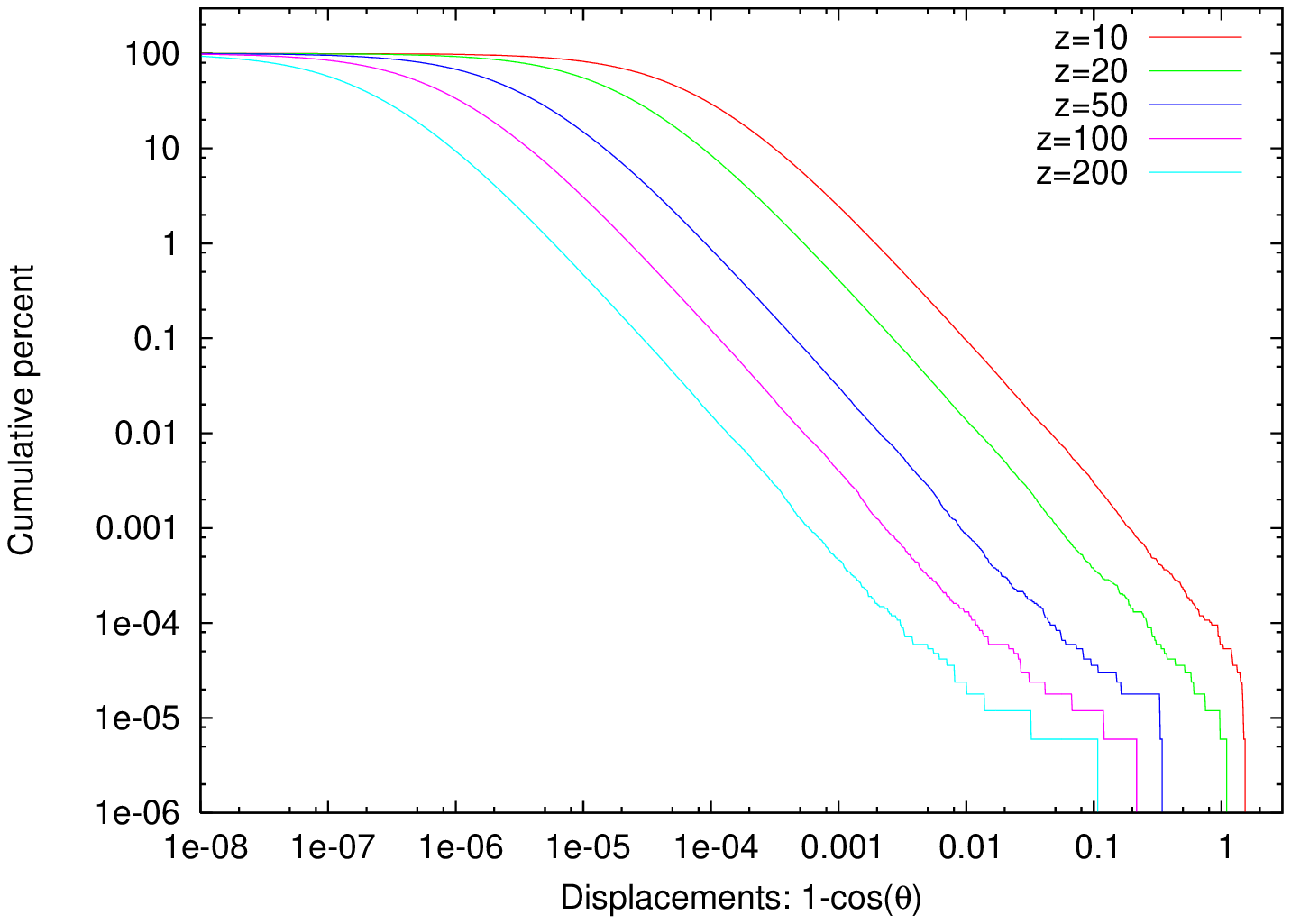}}
\resizebox{2.6in}{!}{\includegraphics[angle=270]{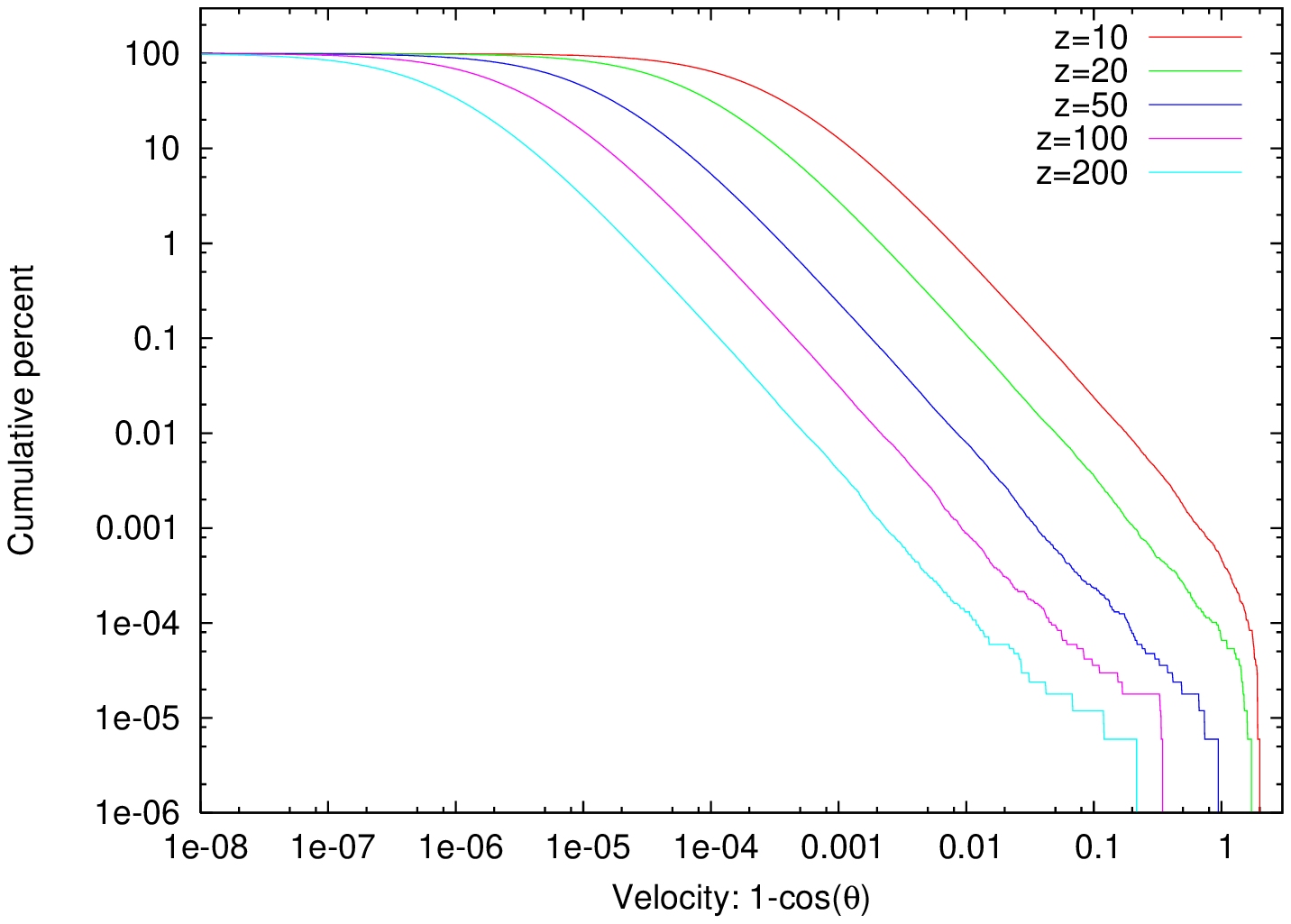}}
\end{center}
\caption{Cumulative distribution of the alignment angles $\cos(\theta)$ between
the Zel'dovich and 2LPT displacement vectors (upper panel) and the
velocity vectors (lower panel) at five different starting redshifts
between $z=200$ and $z=10$. The test was carried out with 256$^3$
particles in a  $(1~h^{-1}$Gpc$)^3$ box. It is clear from these plots
that the curvature in the path is a sub-dominant effect.}
\label{fig:cos}
\end{figure}

An argument as to why 2LPT might be preferable over the Zel'dovich
approximation is that it can capture the displacement curvature,
since it takes into account derivative terms (e.g., \citealt{Bou95}, Fig.~1).
In order to test this hypothesis we measure the distribution of misalignment
angles: $\cos(\theta)$ between the Zel'dovich and 2LPT velocity and
displacement vectors (Figure \ref{fig:cos}).
When starting at high redshift ($z>50$) more than $\sim$99\% of the
particles have paths which differ in direction by less than about $1^\circ$.
Hence the curvature in the path is a small effect for the vast majority of
particles.

A more intuitive understanding of the difference between the Zel'dovich
approximation and 2LPT (in part motivated by Fig.~\ref{fig:vis} of the
velocity field around massive halos in different $z$-start simulations)
is that 2LPT yields a slightly more convergent velocity toward regions
of higher density.  This slightly accelerates massive halo formation
compared to the Zel'dovich approximation, resulting in the change in
the mass function and power spectrum observed.
This picture is supported by the fact that the most massive halos form
about the largest density peaks where one might expect the assumption
of small $\delta$ to hold the least well.

\section{Richardson Extrapolation} \label{app:rich}

Richardson extrapolation is a method to compute the limiting value of a
function that is assumed to have a smooth behavior for small deviations
around the evaluation point.  Suppose we have such a function $f$, then
it is plausible to assume that
\begin{equation}
  f(0+\Delta)=f(0)+c_1\Delta +c_2\Delta^2+ c_3\Delta^3+ \cdots.
\label{smooth}
\end{equation}
For many quantities derived from numerical simulations, it is not
often {\em a priori} obvious what the convergence structure, i.e., the
values of the coefficients, $c_i$, happens to be, even to the extent
of knowing which of the coefficients are zero or
non-zero. Nevertheless, for small enough values of the deviation,
$\Delta$, one can numerically establish the values of the leading
order coefficients.
This allows one to bound the error from a given simulation, and could
even (in principle) allow one to improve estimates for the desired limiting
value $f(0)$ using Richardson extrapolation.

As a simple example, consider the case of non-zero $c_1$ (linear convergence)
for some quantity, say the power spectrum at a given value of $k$, as a
function of the mesh spacing in a PM code.
Then, if we write, for a $256^3$ mesh,
\begin{equation}
  f(2\Delta)\simeq f(0)+2c_1\Delta,\label{256}\\
\end{equation}
for $512^3$ and $1024^3$ meshes we would have,
\begin{eqnarray}
  f(\Delta)&\simeq&f(0)+c_1\Delta,\label{512}\\
  f\left(\frac{\Delta}{2}\right)&\simeq&f(0)+c_1\frac{\Delta}{2} \label{1024},
\end{eqnarray}
where $\Delta$ has been taken to be the mesh spacing for the $512^3$
grid. Eqns.~(\ref{256}) and (\ref{512}) then predict an estimated
value for the 1024$^3$ run
\begin{equation}
  f\left(\frac{\Delta}{2}\right)\simeq\frac 3 2f(\Delta)-\frac 1 2f(2\Delta),
\label{estim_lin}
\end{equation}
which can be used to test whether linear convergence is holding for the
particular range of values of $\Delta$.
If the test is successful, one could then proceed to obtain an estimate for
the continuum prediction ($\Delta=0$) from the 512$^3$ and the 1024$^3$
simulations, via
\begin{equation}
  f(0)\simeq 2 f\left(\frac\Delta 2\right)-f(\Delta).
\label{fzero2}
\end{equation}
We shall require simply that such a prediction differ from our highest
resolution estimate by a negligible amount, to avoid explicit extrapolation.

For the case of quadratic convergence ($c_1=0$, $c_2\neq 0$), the
extrapolation from the 256$^3$ and the 512$^3$ mesh to the 1024$^3$
mesh reads:
\begin{equation}
  f\left(\frac{\Delta^2}{4}\right)\simeq\frac{5}{4}f(\Delta^2)-
  \frac{1}{4}f(4\Delta^2),
\label{estim_quad}
\end{equation}
and the estimate for the continuum from the 512$^3$ simulation and the
1024$^3$ simulation is given by:
\begin{equation}
  f(0)\simeq\frac 4 3 f\left(\frac{\Delta^2}{4}\right)-\frac{1}{3}f(\Delta^2).
\end{equation}
Given 3 simulations one can choose to estimate two non-zero coefficients,
and test the assumed convergence model.  As above, we shall require that
such a prediction differ from our highest resolution estimate by a
negligible amount, to avoid explicit extrapolation.


\begin{thebibliography}{99}

\bibitem[{{Baugh et al.}(1995)}]{baugh}
Baugh, C.M., Gaztanaga, E., \& Estathiou, G., 1995, MNRAS, 274, 1049

\bibitem[{{Benjamin et al.}(2007)}]{Ben07}
Benjamin J., et al., 2007, MNRAS, 381, 702

\bibitem[{{Bernardeau et al.}(2002)}]{Ber02}
Bernardeau F., Colombi S., Gaztanaga E., Scoccimarro R., 2002,
  Phys. Rep. 367, 1

\bibitem[{{Bouchet et al.}(1995)}]{Bou95}
Bouchet, F.R., Colombi, S., Hivon, E., \& Juszkiewicz, R. 1995, A\&A,
296, 575

\bibitem[{{Carlson et al.}(2009)}]{Car09}
Carlson J.W.G., White M., Padmanabhan N., 2009, Phys. Rev. D., in press
  [arxiv:0905.0479]

\bibitem[{{Carroll et al.}(1992)}]{Car92} 
Carroll, S.M., Press, W.H., \& Turner, E.L. 1992, 
Ann.\ Rev.\ Astron.\ Astrophys.\ 30 499

\bibitem[{{Colombi et al.}(2008)}]{Col08}
Colombi, S., Jaffe, A., Novikov, D., \& Pichon, C.,
2008, arXiv:0811.0313

\bibitem[{{Cooray \& Sheth}(2002)}]{HaloModel} 
Cooray, A. \& Sheth, R. 2002, Phys. Rep., 372, 1

\bibitem[{{Crocce et al.}(2006)}]{Cro06}
Crocce, M., Pueblas, S., \& Scoccimarro, R. 2006, MNRAS 373, 369

\bibitem[{{Dunkley et al.}(2008)}]{WMAP5}
Dunkley, J. et al. 2008, arXiv:0803.0586

\bibitem[{{Fu et al.}(2008)}]{Fu08}
Fu L., et al., 2008, A\&A, 479, 9

\bibitem[{{Gottl\"ober \& Klypin}(2008)}]{GotKly08} G\"ottlober, S. \&
  Klypin, A. 2008, in "High Performance Computing in Science and
  Engineering Garching/Munich 2007", Eds. Wagner, S.; Steinmetz, M.;
  Bode, A.; Brehm, M. Springer-Verlag 2008, p.29-43, arXiv:0803.4343

\bibitem[{{Guillet, Teyssier \& Colombi}(2009)}]{GuiTeyCol09}
Guillet T., Teyssier R., Colombi S., 2009, submitted to A\&A
[arxiv:0905.2615] 

\bibitem[{{Habib et al.}(2007)}]{Hab07}
Habib, S., Heitmann, K., Higdon, D., Nakhleh, C., \& Williams B. 2007,
Phys. Rev. D, 76, 083503

\bibitem[{{Hamilton et al.}(1991)}]{Ham91} 
Hamilton, A.J.S., Kumar, P., Lu, E., \& Matthews, A. 1991, ApJ, 
3744, L1

\bibitem[{{Haroz et al.}(2008)}]{Har08} 
Haroz, S., Liu, K.-W., \& Heitmann, K. 2008, IEEE Pacific
Visualization Symposium 2008, 4-7 March 2008, Kyoto, Japan; p.207-14
arxiv:0801.2405 

\bibitem[{{Haroz \& Heitmann}(2008)}]{HarHei08}
Haroz, S. \& Heitmann, K. 2008, IEEE Computer Graphics and
Applications, 28, 37

\bibitem[{{Heitmann et al.}(2005)}]{Hei05}
Heitmann, K., Ricker, P.M., Warren, M.W. \& Habib, S. 2005, ApJS, 28, 160.

\bibitem[{{Heitmann et al.}(2006)}]{Hei06}
Heitmann, K., Higdon, D., Nakhleh, C., \& Habib, S. 2006, ApJ 646, L1

\bibitem[{{Heitmann et al.}(2007)}]{CodeCompare}
Heitmann, K. et al. 2008, Computational Science and Discovery, 1,
015003

\bibitem[{{Heitmann et al.}(2009)}]{Hei09}
Heitmann,~K., Higdon,~D., White,~M., Habib,~S., Williams,~B.J.,
Lawrence,~E, \& Wagner,~C. 2009, ApJ (to appear) 

\bibitem[{{Hilbert et al.}(2008)}]{Hil08} 
Hilbert, S., Hartlap, J., White, S.D.M., \& Schneider, P. 2008, 
arXiv:0809.5035

\bibitem[{{Hockney \& Eastwood}(1989)}]{HocEas89}
Hockney, R.W. \& Eastwood, J.W., ``Computer Simulation Using
Particles'', Publ.: Taylor \& Francis (1989)

\bibitem[{{Hu \& White}(1997)}]{HuWhi97}
Hu, W. \& White, M. 1997, Phys. Rev. D, 56, 596

\bibitem[{{Hu et al.}(1998)}]{HSWZ}
Hu, W., Seljak, U., White, M., \&
  Zaldarriaga, M. 1998, Phys. Rev. D, 57, 3290

\bibitem[{{Huterer \& Takada}(2005)}]{HutTak05}
Huterer, D. \& Takada, M. 2005, Astroparticle Phys. 23, 369

\bibitem[{{Jarvis et al.}(2006)}]{Jar06}
Jarvis,~M., Jain,~B., Bernstein,~G., and Dolney,~D. 2006, ApJ 644, 71

\bibitem[{{Jeong \& Komatsu}(2006)}]{JeoKom06}
Jeong, D. \& Komatsu, E. 2006, ApJ, 651, 619

\bibitem[{{Jing et al.}(2006)}]{Jin06} 
Jing, Y.P., Zhang, P., Lin~W.P., Gao~L., \& Springel, V. 2006, 
ApJ 640, L119

\bibitem[{{Joyce et al.}(2008)}]{Joy08}
Joyce, M., Marcos, B., \& Baertschiger, T., arXiv:0805.1357v1

\bibitem[{{Kilbinger et al.}(2008)}]{kilb08}
Kilbinger, M. et al., arXiv:0810.5129


\bibitem[{{Kravtsov et al.}(1997)}]{Kra97}
Kravtsov, A.V., Klypin, A.A., \& Khokhlov, A. M. 1997, ApJS 111, 73

\bibitem[{{Lawrence et al.}(2009)}]{Law09}
Lawrence,~E., White,~M., Higdon,~D., Wagner,~C., Heitmann, K., Habib,
S., \& Williams, B.J. 2009 (in preparation) `

\bibitem[{{Luki\'c et al.}(2007)}]{Luk07}
Luki\'c, Z., Heitmann, K., Habib, S., Bashinsky, S.,
\& Ricker, P.M. 2007, ApJ 671, 1160

\bibitem[{{Ma}(2007)}]{Ma07}
Ma, Z. 2007, ApJ, 665, 887

\bibitem[{{Massey et al.}(2007)}]{Mas07}
Massey R., et al., 2007, 172, 239

\bibitem[{{Nishimichi et al.}(2008)}]{nish08}
Nishimichi, T. et al., arXiv:0810.0813

\bibitem[{{Peacock}(1999)}]{PeacockBook}
Peacock, J.A., ``Cosmological Physics'', Pub.: Cambridge University
Press (1999)

\bibitem[{{Peacock \& Dodds}(1996)}]{PD96}
Peacock, J.A. \& Dodds, S.J. 1996, MNRAS, 280, L19

\bibitem[{{Rudd et al.}(2008)}]{Rud08}
Rudd, D.H., Zentner, A.R., \& Kravtsov, A.V. 2008, ApJ, 672, 19

\bibitem[{{Schneider et al.}(2008)}]{Sch08}
Schneider, M., Knox, L., Habib, S., Heitmann, K., Higdon, D., \& Nakhleh, C.,
  2008, Phys. Rev. D, 78, 063529

\bibitem[{{Scoccimarro}(1998)}]{Sco98}
Scoccimarro, R. 1998, MNRAS 299, 1097

\bibitem[{{Seljak et al.}(2003)}]{CMBcompare}
Seljak, U., Sugiyama, N., White, M., \& Zaldarriaga, M. 2003,
Phys. Rev. D, 68, 083507

\bibitem[{{Smith et al.}(2003)}]{Smi03} 
Smith, R.E. et al.  [The Virgo Consortium Collaboration], 2003, 
MNRAS 341, 1311

\bibitem[{{Smith et al.}(2007)}]{Smi07}
Smith, R.E., Scoccimarro, R., \& Sheth, R.K. 2007, Phys. Rev. D, 75, 063512

\bibitem[{{Smith et al.}(2008)}]{Smi08}
Smith, R.E., Scoccimarro, R., \& Sheth, R.K. 2008, Phys. Rev. D, 77,
043525

\bibitem[{{Springel}(2005)}]{Spr05}
Springel, V., 2005, MNRAS 364 1105.

\bibitem[{{Tang et al.}(2008)}]{Tan08} 
While there are a large number of ``forecast'' papers, this is one 
of the most recent: Tang,~J., Abdalla, F.B., and Weller, J., 
arXiv:0807.3140 [astro-ph].

\bibitem[{{Takahashi et al.}(2008)}]{Tak08} 
Takahashi, R., Yoshida, N., Matsubara, T., Sugiyama, N., Kayo, I., 
Nishimichi, T., Shirata, A., Taruya, A., Saito, S., Yahata, K., 
\& Suto Y. 2008, MNRAS, 389, 1675

\bibitem[{{White \& Scott}(1996)}]{WhiSco96}
White, M. \& Scott, D. 1996, ApJ, 459, 415

\bibitem[{{White}(2002)}]{Whi02}
White, M. 2002, ApJS, 143, 241

\bibitem[{{White}(2004)}]{Whi04}
White, M. 2004, Astropart. Phys. 22, 211

\bibitem[{{Wong et al.}(2008)}]{Won08}
Wong W.Y., Moss A., Scott D., 2008, MNRAS, 386, 1023

\bibitem[{{Zel'dovich}(1970)}]{Zel70}
Zel'dovich, Y.B. 1970, A\&A, 5, 84

\bibitem[{{Zhan \& Knox}(2004)}]{ZhaKno04}
Zhan, H. \& Knox, L. 2004, ApJ, 616, L75

\bibliographystyle{apj}
\end{thebibliography}
\end{document}